\documentclass[sigconf]{acmart}
\acmConference[ICSE 2024]{46th International Conference on Software Engineering}{April 2024}{Lisbon, Portugal}

\usepackage{amsmath,amsfonts,amsthm}
\usepackage{graphicx}
\usepackage{textcomp}
\usepackage{xcolor}
\usepackage{bbding}
\usepackage[mathscr]{eucal}
\usepackage{multirow}
\usepackage{setspace}
\usepackage{color,framed}
\usepackage{mathrsfs}
\usepackage{tabularx}
\usepackage{float}
\usepackage{url}
\usepackage{enumitem}
\usepackage{balance}
\usepackage{subcaption}
\usepackage{booktabs}
\usepackage{threeparttable}
\usepackage{hyperref}
\usepackage{pifont}
\usepackage{makecell}

\usepackage[most]{tcolorbox}
\definecolor{keys}{RGB}{0,135,0}
\definecolor{gray}{RGB}{216,216,216}

\newcommand{\code}[1]{{\fontfamily{cmtt}\fontseries{m}\fontshape{n}\selectfont\small{#1}}}

\usepackage{xspace}
\newcommand{\sysname}{\textsc{Coca}\xspace}
\newcommand{\Explainer}{\textsc{Coca$_{Exp}$}\xspace}
\newcommand{\Trainer}{\textsc{Coca$_{Tra}$}\xspace}
\newcommand{\Definitiona}{\textsc{Definition 1}\xspace}
\newcommand{\Definitionb}{\textsc{Definition 2}\xspace}
\newcommand{\Definitionc}{\textsc{Definition 3}\xspace}

\AtBeginDocument{%
  \providecommand\BibTeX{{%
    \normalfont B\kern-0.5em{\scshape i\kern-0.25em b}\kern-0.8em\TeX}}}

\copyrightyear{2024}
\acmYear{2024}
\setcopyright{acmlicensed}\acmConference[ICSE '24]{2024 IEEE/ACM 46th International Conference on Software Engineering}{April 14--20, 2024}{Lisbon, Portugal}
\acmBooktitle{2024 IEEE/ACM 46th International Conference on Software Engineering (ICSE '24), April 14--20, 2024, Lisbon, Portugal}
\acmDOI{10.1145/3597503.3639168}
\acmISBN{979-8-4007-0217-4/24/04}

\begin{document}
\newcolumntype{C}[1]{>{\centering\arraybackslash}p{#1}}
\title{\sysname: Improving and Explaining Graph Neural Network-Based Vulnerability Detection Systems}
\thanks{\\
*Xiaobing Sun is the corresponding author.\\
}

\author{Sicong Cao}
\affiliation{%
  \institution{Yangzhou University}
  \city{Yangzhou}
  \country{China}
}
\email{DX120210088@yzu.edu.cn}

\author{Xiaobing Sun}
\authornotemark[1]
\affiliation{%
  \institution{Yangzhou University}
  \city{Yangzhou}
  \country{China}}
\email{xbsun@yzu.edu.cn}

\author{Xiaoxue Wu}
\affiliation{%
  \institution{Yangzhou University}
  \city{Yangzhou}
  \country{China}}
\email{xiaoxuewu@yzu.edu.cn}

\author{David Lo}
\affiliation{%
  \institution{Singapore Management University}
  \country{Singapore}}
\email{davidlo@smu.edu.sg}

\author{Lili Bo}
\affiliation{%
  \institution{Yangzhou University}
  \city{Yangzhou}
  \country{China}}
\affiliation{%
  \institution{Yunnan Key Laboratory of Software Engineering}
  \city{Yunnan}
  \country{China}}
\email{lilibo@yzu.edu.cn}

\author{Bin Li}
\affiliation{%
  \institution{Yangzhou University}
  \city{Yangzhou}
  \country{China}}
\email{lb@yzu.edu.cn}

\author{Wei Liu}
\affiliation{%
  \institution{Yangzhou University}
  \city{Yangzhou}
  \country{China}}
\email{weiliu@yzu.edu.cn}


\begin{abstract}
Recently, Graph Neural Network (GNN)-based vulnerability detection systems have achieved remarkable success. However, the lack of explainability poses a critical challenge to deploy black-box models in security-related domains. For this reason, several approaches have been proposed to explain the decision logic of the detection model by providing a set of crucial statements positively contributing to its predictions. Unfortunately, due to the weakly-robust detection models and suboptimal explanation strategy, they have the danger of revealing spurious correlations and redundancy issue.

In this paper, we propose \sysname, a general framework aiming to 1) enhance the robustness of existing GNN-based vulnerability detection models to avoid spurious explanations; and 2) provide both \emph{concise} and \emph{effective} explanations to reason about the detected vulnerabilities. \sysname consists of two core parts referred to as \emph{Trainer} and \emph{Explainer}. The former aims to train a detection model which is robust to random perturbation based on combinatorial contrastive learning, while the latter builds an explainer to derive crucial code statements that are most decisive to the detected vulnerability via dual-view causal inference as explanations. We apply \sysname over three typical GNN-based vulnerability detectors. Experimental results show that \sysname can effectively mitigate the spurious correlation issue, and provide more useful high-quality explanations.
\end{abstract}

\ccsdesc[500]{Security and privacy~Software security engineering}
\ccsdesc[300]{Software and its engineering~Software maintenance tools}

\keywords{Contrastive Learning, Causal Inference, Explainability}
\maketitle

\section{Introduction}\label{Introduction}
Software vulnerabilities, sometimes called security bugs, are weaknesses in an information system, security procedures, internal controls, or implementations that could be exploited by a threat actor for a variety of malicious ends \cite{johnson2011guide}. As such weaknesses are unavoidable during the design and implementation of the software, and detecting vulnerabilities in the early stages of the software life cycle is critically important \cite{DBLP:journals/infsof/WeiBSLZT23,DBLP:journals/jss/SunYBWWZL23}.

Benefiting from the great success of Deep Learning (DL) in code-centric software engineering tasks, an increasing number of learning-based vulnerability detection approaches \cite{VulDeePecker,SySeVR,TOKEN,BGNN4VD} have been proposed. Compared to conventional approaches \cite{Flawfinder,Checkmarx,DBLP:conf/icse/CaoSWBLWLHOL23,DBLP:conf/sp/CaoHSOZWSBLMLW23} that heavily rely on hand-crafted vulnerability specifications, DL-based approaches focus on constructing complex Neural Network (NN) models to automatically learn implicit vulnerability patterns from source code without human intervention. Recently, inspired by the ability to effectively capture structured semantic information (e.g., control- and data-flow) of source code, Graph Neural Networks (GNN) have been widely adopted by state-of-the-art neural vulnerability detectors \cite{FUNDED,Devign,REVEAL,IVDETECT}.

While demonstrated superior performance, due to the \emph{black-box} nature of NN models, GNN-based approaches fall short in the capability to explain why a given code is predicted as vulnerable \cite{REVEAL,Empirical1}. Such a lack of \textbf{explainability} could hinder their adoption when applied to real-world usage as substitutes for traditional source code analyzers \cite{DBLP:conf/icse/Dam0G18}. To reveal the decision logic behind the binary detection results (vulnerable or not), several approaches have been proposed to provide additional explanatory information. These efforts can be broadly cast into two categories, namely \emph{Global Explainability} and \emph{Local Explainability}. Global explanation approaches leverage the explainability built in specific architectures to understand what features that influence the predictions of the models. A common self-explaining approach is \emph{attention mechanism} \cite{DBLP:conf/nips/VaswaniSPUJGKP17}, which uses weights of attention layers inside the network to determine the importance of each input token. For example, LineVul \cite{LineVul} leverages the self-attention mechanism inside the Transformer architecture to locate vulnerable statements for explanation. However, the global explanation is derived from the training data and thus it may not be accurate for a particular decision of an instance \cite{Pyexplainer}. A more popular approach is local explanation \cite{DBLP:journals/tosem/ZouZXLJY21,DBLP:conf/eurosp/HeJH22}, which adopts perturbation-based mechanisms such as LEMNA \cite{DBLP:conf/ccs/GuoMXSWX18} to provide justifications for individual predictions. The high-level idea behind this approach is to search for important features positively contributing to the model’s prediction by removing or replacing a subset of the features in the input space. IVDetect \cite{IVDETECT} leverages GNNExplainer \cite{DBLP:conf/nips/YingBYZL19} to simplify the target instance to a \emph{minimal} PDG sub-graph consisting of a set of crucial statements along with program dependencies while retaining the initial model prediction.

However, these approaches face two challenges that limit their potentials. Firstly, perturbation-based explanation techniques assume that the detection model is quite robust, i.e., these removed/preserved statements are consistent with the ground truth. Unfortunately, as reported in recent works \cite{DBLP:conf/sigsoft/RabinHA21,DBLP:conf/icse/YangSH022}, simple code edits (e.g., variable renaming) can easily mislead NN models to alter their predictions. As a result, the weak robustness of detection models could lead to spurious explanations even if the vulnerable code is correctly identified. Secondly, most prior methods focus on generating explanations from the perspective of factual reasoning \cite{IVDETECT,DBLP:conf/sigsoft/SunejaZZLM21,DBLP:conf/sigsoft/RabinHA21}, i.e., providing a subset of the input program for which models make the same prediction as they do for the original one. However, such extracted explanations may not be concise enough, covering many redundant statements which are benign but highly relevant to the model's prediction. Therefore, it still requires extensive manpower to analyze and inspect numerous explanation results.

\noindent\textbf{Our Work.} 
To tackle these challenges, we propose \sysname, a novel framework to improve and explain GNN-based vulnerability detection systems via combinatorial \underline{\textbf{Co}}ntrastive learning and dual-view \underline{\textbf{Ca}}usal inference. The key insights underlying our approach include (\ding{182}) enhancing the robustness of existing neural vulnerability detection models to avoid spurious explanations, as well as (\ding{183}) providing both \emph{concise} (preserving a small fragment of code for manual review) and \emph{effective} (covering as many truly vulnerable statements as possible) explanations. To this end, we develop two core parts in \sysname referred to as \emph{Trainer} (abbreviated as \Trainer) and \emph{Explainer} (\Explainer for short).

\noindent\textbf{\sysname Design.}
In the model construction phase, \Trainer first applies six kinds of semantic-preserving transformations as data augmentation operators to generate diverse functionally equivalent variants for each code sample in the dataset. Then, given an off-the-shelf GNN-based vulnerability detection model, \Trainer combines self-supervised with supervised contrastive learning to learn robust feature representations by grouping similar samples while pushing away the dissimilar samples. These robust feature representations will be fed into the classifier to train a robustness-enhanced vulnerability detection model. In the vulnerability explanation phase, we propose a model-agnostic extension based on dual-view causal inference called \Explainer, which integrates factual with counterfactual reasoning to derive crucial code statements that are most decisive to the detected vulnerability as explanations.

\noindent\textbf{Implementation and Evaluations.}
We provide the prototype implementation of \sysname over three state-of-the-art GNN-based vulnerability detection approaches (Devign \cite{Devign}, ReVeal \cite{REVEAL}, and DeepWuKong \cite{DeepWukong}). We extensively evaluate our approach with representative baselines on a large-scale vulnerability dataset comprising well-labeled programs extracted from real-world mainstream projects. Experimental results show that \sysname can effectively improve the vulnerability detection performance of existing NN models and provide high-quality explanations.

\noindent\textbf{Contributions.}
This paper makes the following contributions:
\begin{itemize}[leftmargin=1em]
\item We uncover the spurious correlations and redundancy problems in existing GNN-based explainable vulnerability detectors, and point out that these two issues need to be treated together.

\item We propose \sysname\footnote{\url{https://github.com/CocaVul/Coca}}, a novel framework for improving and explaining GNN-based vulnerability detection systems, in which \Trainer improves the robustness of detection models based on combinatorial contrastive learning to avoid spurious explanations, while \Explainer derives both concise and effective code statements as explanations via dual-view causal inference.

\item We provide prototype implementations of \sysname over three state-of-the-art GNN-based vulnerability detection approaches. The extensive experiments show substantial improvements \sysname brings in terms of the detection capacity and explainability.
\end{itemize}

\section{Background}

\subsection{Problem Formulation}
Instead of exploring new models for more effective vulnerability detection, we focus on a more practical scenario, i.e., explaining the decision logic of off-the-shelf GNN-based vulnerability detection models in a \emph{post-hoc} manner as an input code snippet is predicted as vulnerable. In particular, following the definition in recent works \cite{IVDETECT,DBLP:conf/issta/HuWLPWZ023}, \textbf{our problem} is formalized as:


\textbf{\Definitiona (\emph{Vulnerability Explanation}).}
Given an input program $P = \{s_1, \cdots, s_m\}$ which is detected as vulnerable, the explanation is a set of crucial statements $\{s_i, \cdots, s_j\}$  ($1 \leq i \leq j \leq m$) that are most relevant to the decision of the model, where $s_u$ $(i \leq u \leq j)$ denotes the $u$-th statement in program $P$.

In other words, \textbf{our goal} turns to develop an explanation framework applicable to any GNN-based vulnerability detector to provide not only binary results, but also a few lines of code (i.e., a subset of the input program) as explanatory information, to help security practitioners understand why it is detected as vulnerable.

\begin{figure}[t]
  \centering
  \includegraphics[width=0.9\linewidth]{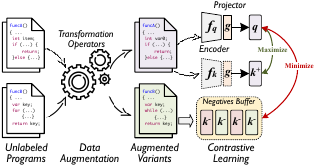}
\caption{Contrastive code representation learning pipeline.}
\label{difference}
\end{figure}

\subsection{Contrastive Learning for Code}\label{SSL4Code}
Due to the limited labeled data in downstream tasks, contrastive learning (CL) has emerged as a promising method for learning better feature representation of code without supervision from labels \cite{ContraBERT,ContraCode,CoLeFunDa,DBLP:conf/acl/DingBPMRC22}. The goal of CL is to maximize the agreement between original data and its positive data (an augmented version of the same sample) while minimizing the agreement between original data and its negative data in the vector space. Figure \ref{difference} presents the typical code-oriented CL pipeline. Unlabeled programs are first transformed into functionally equivalent (FE) variants via data augmentation. In this work, we apply the following six \emph{token-} or \emph{statement-level} augmentation operators introduced by prior works \cite{ContraBERT,ContraCode,DBLP:conf/sigsoft/ChakrabortyADDR22} to construct FE program variants:
\begin{itemize}[leftmargin=1em]
\item \emph{Function/Variable Renaming (FR/VR)}
substitutes the function/variable name with a random token extracted from the vocabulary set constructed on the pre-training dataset.

\item \emph{Operand Swap (OS)}
is to swap the operands of binary logical operations. In particular, the operator will also be changed to make sure the modified expression is the logical equivalent to the one before modification when we swap the operands of a logical operator.

\item \emph{Statement Permutation (SP)}
randomly swaps two lines of statements that have no dependency (e.g., two consecutive declaration statements) on each other in a basic block in a function body.

\item \emph{Loop Exchange (LX)}
replaces for loops with while loops or vice versa.

\item \emph{Block Swap (BS)}
swaps the \code{then} block of a chosen \code{if} statement with the corresponding \code{else} block. We negate the original branching condition to preserve semantic equivalence.

\item \emph{Switch to If (SI)}
replaces a \code{switch} statement in a function body with its equivalent \code{if} statement.
\end{itemize}

Then, these augmented variants are fed into the feature encoder $f_q$ (or $f_k$) with a projection head to produce better global program embeddings via minimizing the contrastive loss. A widely adopted contrastive loss in SE tasks is \emph{Noise Contrastive Estimate} (NCE) \cite{NCE}, which is computed as:
\begin{equation}
\begin{split}
    \mathcal{L}_{NCE} = \frac{1}{|\mathcal{B}|}\sum\limits_{i\in \mathcal{B}}-{\rm log}\frac{{\rm exp}(z_i \cdot z_{j(i)}/\tau)}{\sum\limits_{a\in \mathcal{A}(i)}{\rm exp}(z_i \cdot z_a/\tau)}
\end{split}
\label{NCELoss}
\end{equation}
where $z_i = g(f(\Tilde{x}_i))$ represents the low-dimensional embedding of an arbitrary sample $\Tilde{x}_i$ among augmented variants. $j(i)$ is the index of the other view originating from the same source. $\tau \in \mathcal{R}^+$ is the temperature parameter to scale the loss, and $\mathcal{A}(i) \equiv \mathcal{B} \backslash \{i\}$.

\subsection{Explanation for GNN-based Models}
Although Graph Neural Networks (GNN)-based code models have achieved remarkable success in a variety of SE tasks (e.g., code retrieval \cite{DBLP:conf/sigsoft/ShiYW0Z0ZX22} and automated program repair \cite{DBLP:conf/nips/ChenHLMMTM21}), the lack of explainability creates key barriers to their adoption in practice. Recently, several studies have attempted to explain the decisions of GNNs via \emph{factual reasoning} \cite{DBLP:conf/nips/YingBYZL19,PGExplainer} or \emph{counterfactual reasoning} \cite{CF-GNNExplainer,DBLP:conf/icml/LinLL21}.

\noindent\textbf{Factual Reasoning.}
Factual reasoning-based approaches focus on seeking a sub-graph with a \emph{sufficient} set of edges/features that produce the same prediction as using the whole graph. Formally, given an input graph $\boldsymbol{\mathcal{G}}_k = \{\boldsymbol{\mathcal{V}}_k, \boldsymbol{\mathcal{E}}_k\}$ with its label $\hat{y}_k$ predicted by the trained GNN model, the condition for factual reasoning can be produced as following:
\begin{equation}
\begin{split}
    \underset {\ell \in \boldsymbol{\mathcal{L}}}{\operatorname {arg\,max}}\, P (\ell \mid A_k \odot M_k, X_k \odot F_k) = \hat{y}_k
\end{split}
\label{factualReasoning}
\end{equation}
where $\boldsymbol{\mathcal{L}}$ is the set of graph labels and $\odot$ denotes element-wise multiplication; $M_k \in {\{0, 1\}}^{\boldsymbol{\mathcal{V}}_k\times\boldsymbol{\mathcal{V}}_k}$ represents the edge mask of $\boldsymbol{\mathcal{G}}_k$'s adjacency matrix $A_k \in {\{0, 1\}}^{\boldsymbol{\mathcal{V}}_k\times\boldsymbol{\mathcal{V}}_k}$, while $F_k \in {\{0, 1\}}^{\boldsymbol{\mathcal{V}}_k\times d}$ is the feature mask of $\boldsymbol{\mathcal{G}}_k$'s node feature matrix $X_k \in \mathbb{R}^{\boldsymbol{\mathcal{V}}_k\times d}$. $\boldsymbol{\mathcal{V}}_k$ is the number of nodes in the $k$-th graph and $d$ is the dimension of node features.

\noindent\textbf{Counterfactual Reasoning.}
Counterfactual reasoning-based approaches seek a \emph{necessary} set of edges/features that lead to different predictions once they are removed. Similarly, the condition for counterfactual reasoning can be formulated as:
\begin{equation}
\begin{split}
    \underset {\ell \in \boldsymbol{\mathcal{L}}}{\operatorname {arg\,max}}\, P (\ell \mid A_k - A_k \odot M_k, X_k - X_k \odot F_k) \neq \hat{y}_k
\end{split}
\end{equation}

After optimization, the sub-graph $\boldsymbol{\mathcal{G}}'_k$ will be $A_k \odot M_k$ with the sub-features $X_k \odot F_k$, which is the generated explanations for the prediction of $\boldsymbol{\mathcal{G}}_k$. In our scenario, the extracted sub-graph $\boldsymbol{\mathcal{G}}'_k$ will be further mapped to its corresponding code snippet as explanations for GNN-based vulnerability detectors.

\section{Motivation}\label{Motivation}
\subsection{Special Concerns for DL-based Security Applications}
In contrast to other domains, explanations for security systems should satisfy certain special requirements \cite{DBLP:conf/eurosp/WarneckeAWR20,DBLP:journals/tifs/FanWXLGL21}. In this work, we primarily focus on  two aspects, i.e., \emph{effectiveness} and \emph{conciseness}.

\noindent\textbf{Effectiveness.}
The main goal of an explanation approach is to uncover the decision logic of black-box models. Thus, the vulnerability explainer should be able to capture most relevant features employed by detection models for prediction. For example, given a set of detected vulnerable code, it would be perfect if the provided explanations only cover vulnerability-related context without additional program statements \cite{BTP}.

\textbf{\Definitionb (\emph{Effective Explanations}).}
An explanation result is \emph{effective} if statements which describe the offending execution trace/context of the detected vulnerability are covered.

\noindent\textbf{Conciseness.}
Picking up features/statements highly relevant to the model's prediction is a necessary prerequisite for good explanations. However, it may be difficult and time-consuming for security practitioners to understand and analyze numerous explanation results. Thus, narrowing down the scope of manual review is also important in practice.

\textbf{\Definitionc (\emph{Concise Explanations}).}
An explanation result is \emph{concise} when it only contains a small number of crucial statements sufficient for security experts to understand the root cause of the detected vulnerability.

\subsection{Why Not Fine-Grained Detectors?}
Since the vulnerability explanations are a set of crucial statements derived from the predictions of DL models, an intuitive solution is to construct a fine-grained model to locate vulnerability-related statements, as prior works do \cite{MVD,VulDeeLocator,LineVD,MVD2}. However, the lack of large-scale and human-labeled datasets create key barriers to the adoption of these statement-level approaches in practice. By contrast, we aim to seek a model-independent (or post-hoc) way to provide explanations, instead of replacing them.

\subsection{Why Not Existing Explainers?}
Although the explainability of DL models has been extensively studied in non-security domains \cite{DBLP:conf/cvpr/ZhangWZ18a,DBLP:conf/iccv/FongV17}, we argue that existing explanation approaches face two critical challenges when directly applied to GNN-based vulnerability detection systems.

\noindent\textbf{Weak Robustness.}
As reported in \cite{REVEAL,Empirical1,DBLP:conf/issta/HuWLPWZ023}, existing neural vulnerability detectors focus on picking up dataset nuances for prediction, as opposed to real vulnerability features. Unfortunately, the robustness of most explanation approaches (e.g., LIME \cite{LIME}, SHAP \cite{DBLP:conf/nips/LundbergL17}) are weak, and their explanations for the same sample are easy to be altered due to small perturbations, or even random noise \cite{DBLP:journals/tifs/FanWXLGL21,DBLP:conf/eurosp/WarneckeAWR20}. As a result, explanations built on top of the detection results from such weakly-robust models just reveal spurious correlations, which are hard to be tolerated by security applications.


\noindent\textbf{Hard to Balance Effectiveness and Conciseness.}
Post-hoc approaches mostly explain the predictions made by DL models from the perspective of factual reasoning \cite{IVDETECT,DBLP:conf/ccs/GanzHWR21}, which favors a \emph{sufficient} subset which contains enough information to make the same prediction as they do for the original program. However, such extracted explanations may produce a large number of false alarms, posing a barrier to adoption. What's worse, since the existing post-hoc explanation approaches mainly leverage perturbation-based mechanisms (e.g., LEMNA \cite{DBLP:conf/ccs/GuoMXSWX18}) to track input features that are highly relevant to the model's prediction, the explanation performance will deteriorate further due to the weak robustness of detection models to random perturbations. On the contrary, counterfactual explanations \cite{DBLP:conf/icse/CitoDMC22} contain the most crucial information, which constitutes minimal changes to the input under which the model changes its mind. However, just because of this, they may only cover a small subset of the ground truth.

\subsection{Key Insights Behind Our Design}
In this study, we primarily focus on providing both \emph{effective} and \emph{concise} explanations for security practitioners to gain insights into why a given program was detected as vulnerable. The key insight of \sysname is that the effectiveness and conciseness of explanations can be improved in a two-stage process. This is inspired by the observation that \emph{the robustness of detection models is a necessary prerequisite for effective explanations, while the trade-off between the effectiveness and conciseness mainly depends on the adopted explanation strategy}. Therefore, by employing the two-stage process, the special concerns for effectiveness and conciseness of explanations in GNN-based vulnerability detection systems can be well satisfied.

\begin{figure}[t]
  \centering
  \includegraphics[width=\linewidth]{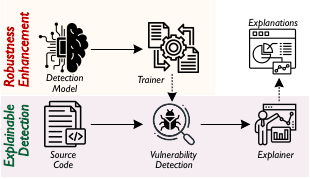}
\caption{The workflow of \sysname.}
\label{Overview}
\end{figure}

\noindent\textbf{Overview.} 
Figure \ref{Overview} presents the workflow of \sysname, including two core components: \emph{Trainer} and \emph{Explainer}. Given a crafted GNN-based vulnerability detection model $\mathcal{M}$, one major difference between our framework and existing approaches lies in the training strategy of the model. Specifically, instead of employing cross-entropy loss, our \emph{Trainer} module leverages combinatorial contrastive loss to train a more robust detector against random perturbations to avoid spurious explanations. Thus, in the vulnerability detection phase, we still transform the input program into graphs and leverage the well-trained model to learn code feature representations for prediction as previous works do. In the explainable detection phase, given a vulnerable code detected by the robustness-enhanced model, we propose a model-agnostic extension, called \emph{Explainer}, to provide security practitioners with both \emph{concise} and \emph{effective} explanations to understand model decisions via dual-view causal inference.

\section{Robustness Enhancement}
\begin{figure}[t]
  \centering
  \includegraphics[width=\linewidth]{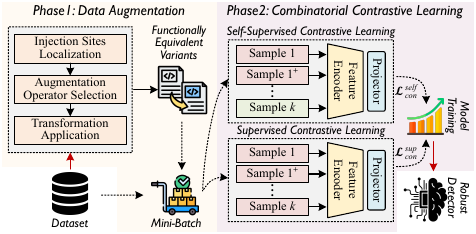}
\caption{The architecture of \Trainer.}
\label{stableTrainer}
\end{figure}
Figure \ref{stableTrainer} depicts the architecture of our \sysname \emph{Trainer} (\Trainer for short). In this stage, we aim to train a neural vulnerability detection model that is robust to random perturbation, which is the core mechanism used in most explainers, to avoid spurious explanations. Specifically, given a crafted detection model $\mathcal{M}$, \Trainer (\ding{182}) augments the vulnerable (or benign) programs in the dataset into a set of functionally equivalent variants via semantics-preserving transformations; and (\ding{183}) leverages combinatorial contrastive learning to force the detection model to focus on critical vulnerability semantics that are consistent between original vulnerable programs and their positive pairs (including perturbed variants and other vulnerable samples), instead of subtle perturbation.

\subsection{Data Augmentation}\label{DataAug}
Inspired by the recent works which adopt obfuscation-based adversarial code as robustness-promoting views \cite{DBLP:journals/corr/abs-2211-11711,DBLP:conf/icml/BielikV20}, our core insight is that the effectiveness of perturbation-based explanation approaches can also benefit from the robustness-enhanced detection models via transformed code because 1) existing perturbation approaches are not suitable for sparse and high-dimensional feature representations of source code \cite{Pyexplainer}; and 2) semantics-preserving code transformations in the discrete token space without changing model predictions can be approximately mapped to the perturbations in the continuous embedding space.

Specifically, to construct functionally equivalent variants, we first perform static analysis to parse each source code $c_i$ into an AST $T_{c_i}$ and traverse it to search for potential injection locations (i.e., AST nodes to which can be applied aforementioned six program transformations $\Phi = \{\phi_1, \phi_2, \cdots, \phi_6\}$). Once an injection location $n_k$ is found, an applicable augmentation operator $\phi_j \in \Phi$ will be randomly selected and applied to get the transformed node $n'_{k}$. We then adapt the context of $n_{k}$ accordingly, and translate it to the FE variant $c'_j$. It is noteworthy that different from synthetic samples \cite{DBLP:conf/icse/NongOPCC23,DBLP:conf/sigsoft/NongOP0C22} used to mitigate the data scarcity issue in classifier training, our transformed FE variants are regarded as augmented views of original samples during contrastive learning to train a robust feature encoder that can capture real vulnerability features. Subsequently, we arrange original code samples along with their FE perturbed variants (positives) as inputs in a mini-batch. In this way, augmented samples originated from one pair are negatively correlated to any sample from other pairs within a mini-batch during contrastive learning.

\subsection{Combinatorial Contrastive Learning}
To train a detection model robust to random perturbations, we borrow the contrastive learning technique to learn better feature representations. Despite the similarity in terms of the high-level design idea \cite{DBLP:conf/sigir/BuiYJ21,DBLP:conf/issta/DingCBPMKR23,ContraCode,ContraBERT}, i.e., pre-training a self-supervised feature-acquisition model over a large unlabeled code database, and performing supervised fine-tuning over labeled dataset to transfer it to a specific downstream SE task, we employ an additional supervised contrastive loss term to effectively leverage label information. Below, we elaborate on each component of our combinatorial contrastive learning with more technical details.

\noindent\textbf{Feature Encoder.}
To extract representations of source code, we employ three typical GNN-based models (Devign \cite{Devign}, ReVeal \cite{REVEAL}, and DeepWuKong \cite{DeepWukong}) as feature encoders $f(\cdot)$. Note that no matter data augmentation or combinatorial contrastive learning are architecture-agnostic, our \Trainer can be easily extended and integrated into other DL-based SE model for robustness enhancement.

\noindent\textbf{Projection Head.}
To improve the representation quality of the feature encoder as well as the convergence of contrastive learning, we add a projection head $g(\cdot)$ consisting of a Multi-Layer Perceptron (MLP) \cite{MLP} with a single hidden layer, to map the embeddings learned by the feature encoder into a low-dimensional latent space to minimize the contrastive loss.

\noindent\textbf{Contrastive Loss.}
Following existing approaches \cite{ContraBERT,ContraCode}, we first employ the NCE loss defined in Eq. (\ref{NCELoss}) as our self-supervised loss function $\mathcal{L}_{con}^{self}$. Specifically, given a set of $N$ randomly sampled unlabeled code samples $\{x_k\}_{k= 1, \cdots, N}$, data augmentation (Section \ref{DataAug}) is applied once to obtain their corresponding FE variants. These samples $\{\Tilde{x}_i\}$, where $\Tilde{x}_{2k-1}$ and $\Tilde{x}_{2k}$ are the original and augmented view of $x_k$, respectively, are then arranged in the mini-batch $\mathcal{B} \equiv \{1, \cdots, 2N\}$ to compute $\mathcal{L}_{con}^{self}$.

In addition, inspired by a recent finding \cite{DBLP:journals/corr/abs-2211-11711} that the robustness enhanced in the self-supervised pre-training phase may no longer hold after supervised fine-tuning, we also adopt the \emph{Supervised Contrastive} (SupCon) loss \cite{SupCon} during the training process because the use of label information encourages the feature encoder to closely aligns all samples from the same class in the latent space to learn more robust (in terms of original samples and FE variants) and accurate (in terms of samples with the same label) cluster representations. Formally, the SupCon loss $\mathcal{L}_{con}^{sup}$ is written as:
\begin{equation}
\begin{split}
    \mathcal{L}_{con}^{sup} = \frac{1}{|\mathcal{B}^l|}\sum\limits_{i\in \mathcal{B}^l}\frac{-1}{|\mathcal{Q}(i)|}\sum\limits_{q\in \mathcal{Q}(i)}{\rm log}\frac{{\rm exp}(z_i \cdot z_q/\tau)}{\sum\limits_{a\in \mathcal{A}(i)}{\rm exp}(z_i \cdot z_a/\tau)}
\end{split}
\end{equation}
where $\mathcal{B}^l$ corresponds to the subset (known vulnerable or benign code) of $\mathcal{B}$, and $\mathcal{Q}(i) \equiv \{q \in \mathcal{A}(i) : \Tilde{y}_q = \Tilde{y}_i\}$ is the set of indices of all other \emph{positives} that hold the same label as $\Tilde{x}_i$ in $\mathcal{B}$. $1/|\mathcal{Q}(i)|$ is the positive normalization factor which serves to remove bias present in multiple positives samples and preserve the summation over negatives in the denominator to increase performance.

Finally, the total loss used to train a robust feature encoder over the batch is defined as:
\begin{equation}
\begin{split}
    \mathcal{L}_{total} = (1-\lambda)\mathcal{L}_{con}^{self} + \lambda\mathcal{L}_{con}^{sup}
\end{split}
\end{equation}
where $\lambda$ is a weight coefficient to balance the two loss terms.

At the end of combinatorial contrastive learning, the projection head $g(\cdot)$ will be discarded and the well-trained feature encoder $f(\cdot)$ is frozen (i.e., containing exactly the same number of parameters when applied to specific downstream tasks) to produce the vector representation of a program for vulnerability detection.

\section{Explainable Detection}
The explainable detection stage aims to (\ding{182}) train a classifier on top of the robust feature encoder for vulnerability detection; and (\ding{183}) build a explainer to derive crucial statements as explanations.

\subsection{Vulnerability Detection}
The goal of this task is to train a binary classifier able to accurately predict the probability that a given function is vulnerable or not. In particular, given a popular GNN-based vulnerability detection model, we only replace its feature encoder with the more robust one\footnote{Note that the feature encoder is fixed during the whole training phase.} (sharing the same NN architecture) which is pre-trained by \Trainer. Thus, any coarse-grained (function- or slice-level) vulnerability detector, which receives structural graph representations of source code (in which code tokens/statements are nodes while semantic relations between nodes are edges) as inputs and employs an off-the-shelf or crafted GNN as its feature encoder for vulnerability feature learning, can be easily integrated into our framework. For example, when ReVeal \cite{REVEAL} is selected as the target detection model, labeled code snippets are parsed into Code Property Graphs (CPGs) \cite{DBLP:conf/sp/YamaguchiGAR14} and fed into the robust feature encoder $f(\cdot)$ (a vanilla GGNN \cite{GGNN}) pre-trained by \Trainer to produce corresponding vector representations. Then, these representations and their labels are used to train a built-in classifier (a convolutional layer with maxpooling) with the triplet loss function.

In the inference phase, given an input program, the vulnerability detector first performs static analysis to extract its graph representation and maps it as a single vector representation using the pre-trained feature encoder. Then, the program representation will be fed into the trained classifier for prediction.

\subsection{Vulnerability Explanation}
To derive explanations on why the detection model has decide on the vulnerability, we propose a model-agnostic extension based on the detection results, referred to as \emph{Explainer} (\Explainer for short).

\noindent\textbf{Overview.}
Similar to the most related work IVDetect \cite{IVDETECT}, \Explainer aims to find a sub-graph $\mathcal{G}'_k$, which covers the key nodes (tokens/statements) and edges (program dependencies) that are most decisive to the prediction label, from the graph representation $\mathcal{G}_k$ of the detected vulnerable code $k$. The main difference lies in that we aim to seek both concise and effective explanations. Hence, we build \Explainer based on a dual-view causal inference framework \cite{CF2} which integrates factual with counterfactual reasoning to make a trade-off between conciseness and effectiveness. Formally, the extraction of $\mathcal{G}'_k$ can be formulated as:
\begin{equation}
\begin{aligned}
    &{\rm minimize} \ C(M_k, F_k) \\
    &{\rm s.t.}, S_f(M_k, F_k) > P(\hat{y}_{k, s} \mid A_k \odot M_k, X_k \odot F_k), \\
    &S_c(M_k ,F_k) > - P(\hat{y}_{k, s} \mid A_k - A_k \odot M_k, X_k - X_k \odot F_k) \\
\end{aligned}
\label{optimization}
\end{equation}
where the objective part $C(M_k, F_k)$ measures how concise the explanation is. It can be defined as the number of edges/features used to generate the explanation sub-graph $\mathcal{G}'_k$, and computed by $C(M_k, F_k) = \Vert M_k \Vert_0 + \Vert F_k \Vert_0$, in which $\Vert M_k \Vert_0$ (/$\Vert F_k \Vert_0$) represents the number of 1’s in the binary edge mask $M_k$ (/feature mask $F_k$) metrices. The constraint part $S_f(M_k, F_k)$ (/$S_c(M_k ,F_k)$) reflects whether the factual (/counterfactual) explanation is effective enough. Formally, the factual explanation strength $S_f(M_k, F_k)$ is consistent with the condition for factual reasoning, i.e., $S_f(M_k, F_k) = P(\hat{y}_k \mid A_k \odot M_k, X_k \odot F_k)$. Similarly, the counterfactual explanation strength $S_c(M_k ,F_k)$ is calculated as $S_c(M_k, F_k) = -P(\hat{y}_k \mid A_k - A_k \odot M_k, X_k - X_k \odot F_k)$. $\hat{y}_{k, s}$ is the label other than $\hat{y}_k$ that has the largest probability score predicted by the GNN-based detection model.

To solve such a constrained optimization problem, we follow \cite{CF2}, which optimizes the objective part by relaxing $M_k$ and $F_k$ to real values $M_k^* \in \mathbb{R}^{\boldsymbol{\mathcal{V}}_k\times \boldsymbol{\mathcal{V}}_k}$ and $F_k^* \in \mathbb{R}^{\boldsymbol{\mathcal{V}}_k\times d}$, and using 1-norm to ensure the sparsity of $M_k^*$ and $F_k^*$. For the constraint part, we relax it as pairwise contrastive loss $\mathcal{L}_f$ and $\mathcal{L}_c$:
\begin{equation}
\begin{aligned}
    \mathcal{L}_f = \ &{\rm ReLU}(\frac{1}{2} -S_f(M_k^*, F_k^*) \\
          &+ P(\hat{y}_{k, s} \mid A_k \odot M_k^*, X_k \odot F_k^*)) \\
    \mathcal{L}_c = \ &{\rm ReLU}(\frac{1}{2} - S_c(M_k^*, F_k^*) \\
          &- P(\hat{y}_{k, s} \mid A_k - A_k \odot M_k^*, X_k - X_k \odot F_k^*))
\end{aligned}
\end{equation}

After that, the explanation sub-graph $\mathcal{G}'_k=(A_k \odot M_k^*, X_k \odot F_k^*)$ is generated by:
\begin{equation}
\begin{aligned}
    &{\rm minimize} \ \Vert M_k^* \Vert_1 + \Vert F_k^* \Vert_1 + \alpha \mathcal{L}_f + (1- \alpha)\mathcal{L}_c
\end{aligned}
\label{ralaxminize}
\end{equation}
Where $\alpha$ controls the trade-off between the strength of factual and counterfactual reasoning. By increasing/deceasing $\alpha$, the generated explanations will focus more on the effectiveness/conciseness.

\section{Experiments}

\subsection{Research Questions}
\noindent In this paper, we seek to answer the following RQs:

\noindent\textbf{RQ1 (Detection Performance):}
How effective are existing GNN-based approaches enhanced via \sysname on vulnerability detection?

The disconnection between the learned features versus the actual cause of the vulnerabilities has raised the concerns regarding the effectiveness of DL-based detection models. Thus, we investigate whether the enhanced GNN-based vulnerability detectors outperform their original ones in terms of detection accuracy and the ability to capture real vulnerability features after robustness enhancement.

\noindent\textbf{RQ2 (Explanation Performance):}
Is \sysname more concise and effective than state-of-the-art baselines when applied to generate explanations for GNN-based vulnerability detectors?

We argue that generating corresponding explanations for detection results is just the first step and the quality evaluation of them is also important. With this motivation, we evaluate the performance of \sysname in generating  concise and effective explanations.

\noindent\textbf{RQ3 (Ablation Study):}
How do various factors affect the overall performance of \sysname?

We perform sensitivity analysis to understand the influence of different components of \sysname, including the impact of (\textbf{RQ3a}) combinatorial contrastive learning, and (\textbf{RQ3b}) dual-view causal inference.

\subsection{Datasets}
\begin{table}[t]
 \caption{The statistics of datasets.}
  \centering
  \renewcommand\arraystretch{0.75}
  \begin{tabular}{c|rrrr}
    \toprule
    \textbf{Dataset} &\textbf{\# Vul} &\textbf{\# Non-vul} &\textbf{\# Total} &\textbf{\% Ratio}\\
    \midrule
    Devign   & 11,888   & 14,149   & 26,037   & 45.66   \\
    ReVeal   & 1,664    & 16,505   & 18,169   & 9.16   \\
    Big-Vul  & 11,823   & 253,096  & 264,919  & 4.46   \\
    CrossVul & 6,884    & 127,242  & 134,126  & 5.13   \\
    CVEFixes & 8,932    & 159,157  & 168,089  & 5.31   \\
    \midrule
    \textbf{Merged} & \textbf{29,844} & \textbf{305,827} & \textbf{335,671} & \textbf{8.89}\\
    \bottomrule
  \end{tabular}
  \label{dataset}
\end{table}

Since the detection capability of DL-based models benefits from large-scale and high-quality datasets, we built our evaluation benchmark by merging five reliable human-labeled datasets collected from real-world projects, including \textbf{Devign} \cite{Devign}, \textbf{ReVeal} \cite{REVEAL}, \textbf{Big-Vul} \cite{FAN}, \textbf{CrossVul} \cite{CrossVul}, and \textbf{CVEFixes} \cite{CVEfixes}. Detailed statistics for each of the five datasets is shown in Table \ref{dataset}. Column 2 and Column 3 are the number of vulnerable and non-vulnerable functions, respectively. Column 4 indicates the total number of functions in each dataset. Column 5 denotes the ratio of vulnerable functions in each dataset. Note that, for two multi-language datasets \textbf{CrossVul} and \textbf{CVEFixes}, we only preserve code samples written in C/C++ in our experiments to unify the whole dataset. In total, our merged dataset contains 335,671 functions, of which 29,844 (8.89\%) are vulnerable.

\subsection{\sysname Implementation}
For \Trainer, we parsed all the code snippets in our merged dataset into ASTs using \code{tree-sitter}\footnote{\url{https://tree-sitter.github.io/tree-sitter/}} and performed the transformation based on augmentation operators described in Section \ref{DataAug} to generate perturbed variants. We applied all six transformations with an equal probability of 0.5, which leads us to an average of three transformations per program. In contrastive learning, any GNN-based detection model can be served as an feature encoder in our framework and trained on an Ubuntu 18.04 server with 2 NVIDIA Tesla V100 GPU. Following standard practice in contrastive code representation learning \cite{CoLeFunDa,ContraCode}, we set the size of the projection head to 128, and used Adam \cite{ADAM} for optimizing with 256 batch size and 1$e$-5 learning rate. The temperature parameter $\tau$ of contrastive loss is set to 0.07. For feature encoder training, we randomly sampled a subset (50\%) of vulnerable and benign samples from the merged dataset, respectively, to construct $\mathcal{B}$, and the remaining samples are regarded as the unlabeled data $\mathcal{B}^l$. The feature encoder and classifier of each detection model were trained with 100 maximum epochs and early stopping. For \Explainer, we set $\alpha$ to 0.5 to balance factual and counterfactual reasoning.

\section{Experimental Results}
\subsection{RQ1: Detection Performance}
\begin{table}[t]\small
 \caption{Evaluation results on vulnerability detection in percentage compared with GNN-based baselines.}
  \centering
  \setlength\tabcolsep{2.5pt}
  \renewcommand\arraystretch{0.75}
  \begin{tabular}{c|c|l|cccc}
    \toprule
    \textbf{Config} & \textbf{Loss} & \textbf{Approach}     & \textbf{Acc}    & \textbf{Pre}    & \textbf{Rec}      & \textbf{F1} \\
    \midrule
    \multirow{3}*{Default} &\multirow{3}*{CE} &Devign      & \textbf{89.74} & 32.59 & 31.40 & 31.98   \\
    &&ReVeal      & 86.05 & 31.43 & 38.45 & 34.59   \\
    &&DeepWuKong  & 87.21 & 28.55 & 26.04 & 27.24   \\
    \midrule[0.05em]
    \multirow{9}*{\Trainer} &\multirow{3}*{Ours} &Devign      & 88.15 & 34.68 & 37.12 & 35.86   \\
    &&ReVeal      & 87.42 & \textbf{35.96} & \textbf{40.61} & \textbf{38.14}   \\
    &&DeepWuKong  & 88.30 & 30.07 & 34.79 & 32.26   \\
    \cmidrule[0.05em]{2-7}
    &\multirow{3}*{InfoNCE} & Devign      & 86.33 & 28.38 & 30.11 & 29.22   \\
    &&ReVeal      & 84.95 & 29.64 & 34.27 & 31.78   \\
    &&DeepWuKong  & 86.20 & 25.99 & 24.83 & 25.40   \\
    \cmidrule[0.05em]{2-7}
    & \multirow{3}*{NCE} &Devign      & 83.97 & 26.15 & 27.69 & 26.90   \\
    &&ReVeal      & 81.52 & 26.73 & 31.76 & 29.03   \\
    &&DeepWuKong  & 83.06 & 22.40 & 21.46 & 21.92   \\
    \bottomrule
  \end{tabular}
  \label{Detection results}
\end{table}

\underline{\emph{Baselines.}}
We consider three state-of-the-art GNN-based vulnerability detectors: 1) \textbf{Devign} \cite{Devign} models programs as graphs and adopts GGNN \cite{GGNN} to capture structured vulnerability semantics; 2) \textbf{ReVeal} \cite{REVEAL} adopts graph embedding with triplet loss function to learn class-separation vulnerability features; and 3) \textbf{DeepWuKong} \cite{DeepWukong} leverages GCN to learn both unstructured and structured vulnerability information at the slice-level.

\underline{\emph{Evaluation Metrics.}} We apply four widely used metrics \cite{DBLP:journals/csur/PendletonGCX17}, including \textbf{Accuracy (Acc)}, \textbf{Precision (Pre)}, \textbf{Recall (Rec)}, and \textbf{F1-score (F1)}, for evaluation.

\underline{\emph{Experiment Setup.}}
For the open-source approaches (ReVeal and DeepWuKong), we directly use their official implementations. For Devign, which is not publicly available, we re-implemented it by strictly following its methods elaborated in the original paper. In addition, to integrate these approaches into \sysname, we also employ \code{tree-sitter} to uniformly parse input programs into their expected graph representations (e.g., PDG, CPG). We randomly split the benchmark into 80\%-10\%-10\% for training, validation, and testing. For each approach, we repeated the experiment 10 times to address the impact of randomness \cite{DBLP:conf/icse/ArcuriB11,DBLP:journals/tse/Tantithamthavorn17}.

\underline{\emph{Results.}}
Table \ref{Detection results} summarizes the experimental results of all the studied baselines and their corresponding variants enhanced by \Trainer on vulnerability detection. Column "Config" presents the configuration of GNN-based vulnerability detectors, i.e., constructing detection models with default implementations (supervised learning with CE loss) or \Trainer (contrastive learning with NCE and SupCon loss). Overall, the average improvements of robustness-enhanced models over their default ones are positive, ranging from 5.32\% (DeepWuKong) to 14.41\% (ReVeal) on Precision, from 5.62\% (ReVeal) to 33.60\% (DeepWuKong) on Recall, and from 10.26\% (ReVeal) to 18.43\% (DeepWuKong) on F1, respectively. In addition, \Trainer (ReVeal) achieves the overall best performance, with an Accuracy of 87.42\%, the Precision of 35.96\%, the Recall of 40.61\%, and the F1 of 38.14\%.

\begin{figure}[t]
\centering
\begin{subfigure}{0.49\linewidth}
\centering
\includegraphics[width=\linewidth]{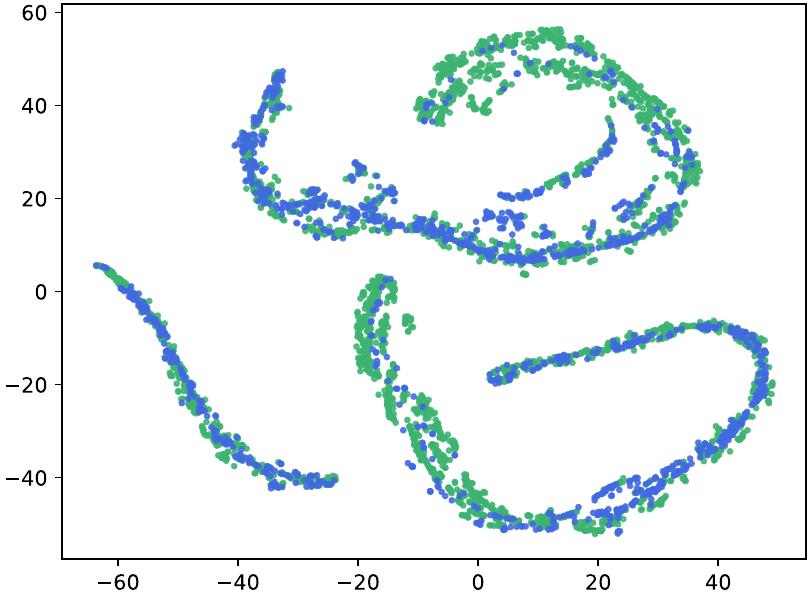}
\caption{DeepWuKong (Default)}
\end{subfigure}
\begin{subfigure}{0.49\linewidth}
\centering
\includegraphics[width=\linewidth]{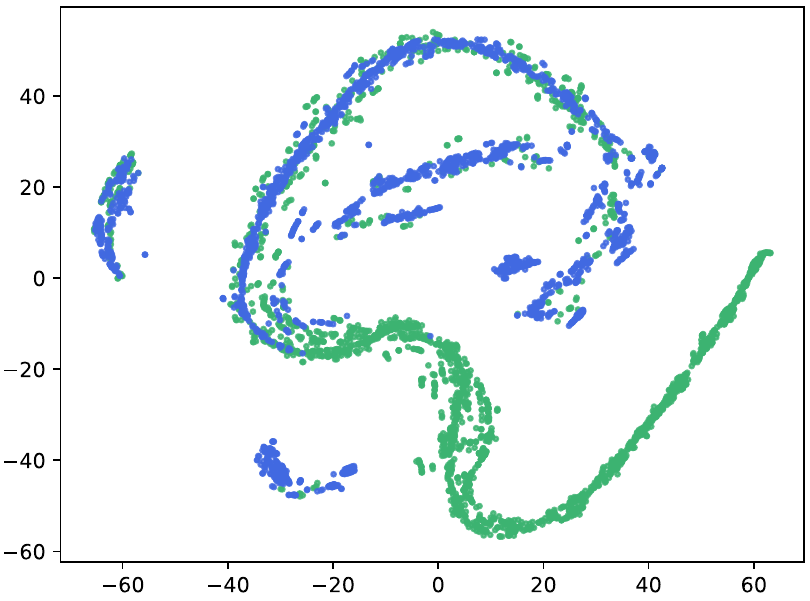}
\caption{DeepWuKong (\Trainer)}
\end{subfigure}
\caption{\label{T-SNE}Visualizations of feature representations learned by DeepWuKong trained with/without \Trainer.}
\end{figure}

All these results demonstrate the effectiveness of \Trainer in improving the vulnerability detection performance of existing GNN-based code models. It indicates that incorporating structurally perturbed samples (e.g., statement permutation, loop exchange) into contrastive learning is beneficial for the graph-based model to focus on security-critical structural semantics rather than noise information. Taking the greatest improved model DeepWuKong as an example, as shown in the visualizations in Figure \ref{T-SNE}, the feature representations learned by \Trainer(DeepWuKong) are more class-discriminative compared to the ones learned with default cross-entropy loss. We attribute such improvements to robustness-enhanced models truly capturing discriminative vulnerability patterns from the comparison between vulnerable samples and perturbed/benign variants. 

\begin{tcolorbox}[left=1pt,right=1pt,top=1pt,bottom=1pt,boxrule=0.7pt,enhanced,drop fuzzy shadow]
\textbf{Answer to RQ1: }\Trainer comprehensively improves the performance of existing GNN-based vulnerability detectors in terms of all evaluation metrics. We attribute the improvements to the robustness-enhanced models truly picking up real vulnerability features for prediction.
\end{tcolorbox}

\subsection{RQ2: Explanation Performance}

\underline{\emph{Baselines.}}
We adopt three recent vulnerability explanation approaches as baselines: 1) \textbf{IVDetect} \cite{IVDETECT} leverages GNNExplainer \cite{DBLP:conf/nips/YingBYZL19} to produce the key program dependence sub-graph (i.e., a list of crucial statements closely related to the detected vulnerability) that affect the decision of the model as explanations; 2) \textbf{P2IM} \cite{DBLP:conf/sigsoft/SunejaZZLM21} borrows \emph{Delta Debugging} \cite{DBLP:journals/tse/ZellerH02} to reduce a program sample to a minimal snippet which a model needs to arrive at and stick to its original vulnerable prediction to uncover the model's detection logic; and 3) \textbf{mVulPreter} \cite{mVulPreter} combines the attention weight with the vulnerability probability outputted by the multi-granularity detector to compute the importance score for each code slice.

\underline{\emph{Evaluation Metrics.}} As we described in Section \ref{Motivation}, an ideal explanation should cover as many truly vulnerable statements (in terms of \emph{effectiveness}) as possible within a limited scope (in terms of \emph{conciseness}). Thus, we use the fine-grained \textbf{Vulnerability-Triggering Paths (VTP)} \cite{ContraFlow,BTP} metrics to evaluate the quality of explanations, which are formally defined as follows:

\begin{itemize}[leftmargin=1em]
\item \textbf{\emph{Mean Statement Precision} (MSP):} $MSP = \frac{1}{N}\sum_{i=1}^NSP_i$ where $SP_i = |S_e \cap S_p|/|S_e|$ stands for the proportion of contextual statements truly related to the detected vulnerability sample $i$ in the explanations.

\item \textbf{\emph{Mean Statement Recall} (MSR):} $MSR = \frac{1}{N}\sum_{i=1}^NSR_i$, where $SR_i = |S_e \cap S_p|/|S_p|$ denotes that how many contextual statements in the triggering path of the detected vulnerability sample $i$ can be covered in explanations.

\item \textbf{\emph{Mean Intersection over Union} (MIoU):} $MIoU = \frac{1}{N}\sum_{i=1}^NIoU_i$, where $IoU_i = |S_e \cap S_p|/|S_e \cup S_p|$ reflects the degree of overlap between the explanatory statements and the contextual statements on the vulnerability-triggering path.
\end{itemize}
Here, $S_e$ denotes the set of explanatory statements provided by explainers, while $S_p$ denotes the set of labeled vulnerability-contexts (ground truth) in the dataset. $|\cdot|$ represents the size of a set.

\underline{\emph{Experiment Setup.}}
We still employ three aforementioned GNN-basd vulnerability detectors (with default/robustness-enhanced configurations) to provide prediction labels for IVDetect\footnote{Since IVDetect implements its own vulnerability detector based on FA-GCN, we do not use other detection models as alternatives.}, P2IM, and \Explainer. For mVulPreter, we follow the official implementation to produce explanations because its detection and explanation module is highly coupled. For two baselines (IVDetect and mVulPreter) which require a human-selected $k$ value to decide the size of the explanations, we follow \cite{IVDETECT,mVulPreter} to narrow down the scope of candidate statements to 5, while the size of explanations produced by our approach and P2IM are automatically decided by themselves via optimization. Following \cite{DBLP:conf/sigsoft/SunejaZZLM21,ContraFlow}, we evaluate these explanation approaches on another vulnerability dataset \textbf{D2A} \cite{D2A} because it is labeled with clearly annotated vulnerability-contexts which are more reliable than other \emph{diff}-based ground truths \cite{BTP}. We randomly select 10,000 vulnerable samples which can be correctly detected from the D2A dataset to calculate the VTP metrics.

\begin{table}[t]
 \caption{Evaluation results on vulnerability explanation in percentage compared with explainable vulnerability detection baselines.}
  \centering
  \renewcommand\arraystretch{0.75}
  \begin{tabular}{c|l|ccc}
    \toprule
    \textbf{Config} & \textbf{Approach}     & \textbf{MSP}    & \textbf{MSR}    & \textbf{MIoU}\\
    \midrule
    \multirow{8}*{\rotatebox{90}{Default}} & mVulPreter  & 25.86 & 29.01 & 22.88 \\
    & IVDetect                  & 32.54 & 23.79 & 17.06 \\
    & P2IM (Devign)             & 27.99 & 43.85 & 22.56 \\  
    & P2IM (ReVeal)             & 31.04 & 46.10 & 28.94 \\
    & P2IM (DeepWuKong)         & 26.57 & 38.12 & 23.11 \\
    & \Explainer (Devign)       & 33.84 & 44.06 & 30.89 \\
    & \Explainer (ReVeal)       & 35.61 & 52.94 & 34.36 \\
    & \Explainer (DeepWuKong)   & 29.77 & 40.16 & 25.83 \\
    \midrule
    \multirow{7}*{\rotatebox{90}{\Trainer}} & IVDetect   & 39.81 & 31.64 & 25.19 \\
    & P2IM (Devign)             & 33.01 & 48.33 & 29.27 \\
    & P2IM (ReVeal)             & 40.62 & 55.73 & 36.29 \\
    & P2IM (DeepWuKong)         & 32.97 & 44.85 & 28.10 \\
    & \Explainer (Devign)       & 43.61 & 52.98 & 39.64 \\
    & \Explainer (ReVeal)       & \textbf{49.52} & \textbf{58.39} & \textbf{44.97} \\
    & \Explainer (DeepWuKong)   & 40.33 & 47.61 & 34.22 \\
    \bottomrule
  \end{tabular}
  \label{Explanation results}
\end{table}

\underline{\emph{Results.}}
Table \ref{Explanation results} shows the performance comparison of \Explainer with respect to state-of-the-art explanation approaches. As can be seen, based on the predictions of popular graph-based vulnerability detectors (with default implementations), \Explainer substantially outperforms all the compared explanation techniques on all metrics. Taking the best comparison baseline P2IM (ReVeal) as an example, \Explainer (ReVeal) outperforms it by 14.72\% in MSP, 14.84\% in MSR, and 18.73\% in MIoU, respectively.

In addition, although there is still a certain gap from our best-performing \Explainer, we find that the performance of each explanation baseline can be improved to varying degrees when applied to robustness-enhanced detection models. The main reason leading to this result is that the more robust feature representations gained by contrastive learning can better reflect the potential vulnerable behaviour of programs and boost vulnerability semantic comprehension. Among them, \Explainer (ReVeal) yields the best explanation performances on all metrics (especially MIoU), demonstrating that our dual-view causal inference makes a great trade-off between the effectiveness (covering as many truly vulnerable statements as possible) and conciseness (limiting the number of candidates for manual review) of explanations. Meanwhile, we notice that the attention-based explainer mVulPreter performs extremely poorly on the vulnerability explanation task. The reason is that attention weights are derived from the training data \cite{CCzhu}. Thus, it may not be accurate for a particular decision of an instance. On the contrary, \Trainer and the other two vulnerability explainers (IVDetect and P2IM) construct an additional explanation model for an individual instance in a model-independent manner to provide explanatory information, effectively avoiding the decision bias.

\begin{figure}[t]
  \centering
  \includegraphics[width=\linewidth]{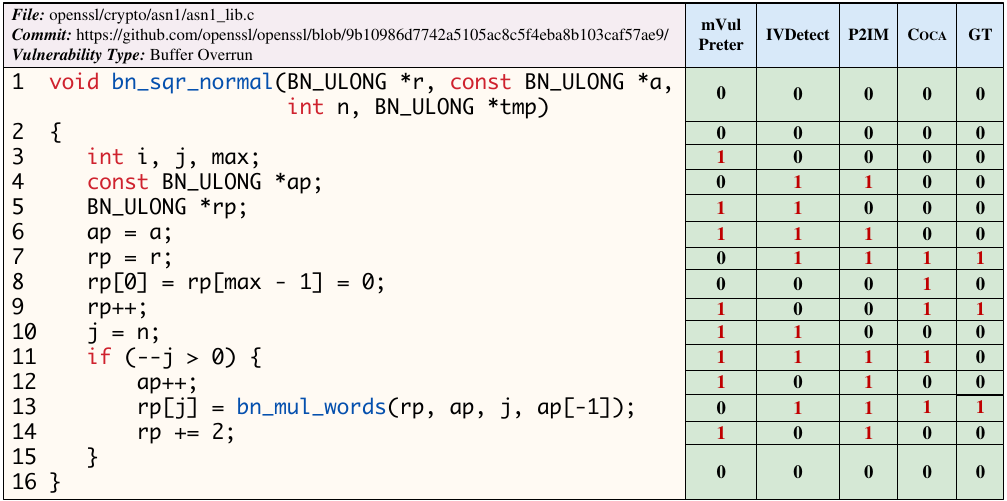}
\caption{Qualitative study of our \sysname vs. baselines.}
\label{CaseStudy}
\end{figure}

To gain a more intuitive understanding of how effective and concise our generated explanations are, we perform a qualitative study to evaluate the quality of explanations generated by \sysname and other explainers. To ensure the fairness, the explanations provided by two explainers (P2IM and \sysname) not dependent on specific detectors, are generated from the robustness-enhanced model ReVeal. Figure \ref{CaseStudy} shows a correctly detected vulnerable function in the D2A dataset. Column "GT" denotes the ground truth. It contains a \emph{Buffer Overrun} vulnerability in \code{rp} (at line 13) when calling the function \code{bn\_mul\_words()}. Statements at line 7 and line 9 are its corresponding vulnerability-contexts annotated by D2A. Overall, \emph{all} three vulnerable statements are covered by \sysname, while mVulPreter, IVDetect, and P2IM could only report one, two, and two of them, respectively. Furthermore, in terms of the conciseness, three of five explanatory statements provided by \sysname are true positives, with a recall of 60\%. By contrast, 87.5\%, 71.43\%, and 71.43\% statements included in the explanations of mVulPreter, IVDetect, and P2IM are false positives. Therefore, \sysname can provide as many truly vulnerable statements as possible within a limited scope to help security practitioners understand the detection results provided by GNN-based vulnerability detection systems.

\begin{tcolorbox}[left=1pt,right=1pt,top=1pt,bottom=1pt,boxrule=0.7pt,enhanced,drop fuzzy shadow]
\textbf{Answer to RQ2: }\Explainer is superior to the state-of-the-art explainers in terms of the effectiveness and conciseness. When applied to the best detection model \Trainer (ReVeal), \Explainer improves MSP, MSR, and MIoU over the best-performing baseline P2IM by 21.91\%, 4.77\%, and 23.92\%, respectively.
\end{tcolorbox}

\subsection{RQ3: Ablation Study}

\underline{\emph{Baselines.}}
For \textbf{RQ3a}, we compare \Trainer with three representative loss functions: 1) \textbf{NCE} \cite{NCE} frames contrastive learning as a self-supervised binary classification problem, which predicts whether a data point came from the noise distribution or the true data distribution; 2) \textbf{InfoNCE} \cite{InfoNCE} generalizes NCE loss by computing the probability of selecting the positive sample across a batch and a queue of negatives; and 3) \textbf{Cross-Entropy (CE)}, the most widely used supervised loss for deep classification models. For \textbf{RQ3b}, we compare \Explainer with the following GNN-specfic explanation approaches: 1) \textbf{GNNExplainer} \cite{DBLP:conf/nips/YingBYZL19} selects a discriminative sub-graph that retains important edges/node features via maximizing the mutual information of a prediction; 2) \textbf{PGExplainer} \cite{PGExplainer} uses an explanation network on a universal embedding of the graph edges to provide explanations for multiple instances; and 3) \textbf{CF-GNNExplainer} extends GNNExplainer by generating explanations based on counterfactual reasoning.

\underline{\emph{Experiment Setup.}}
To answer RQ3a, we built three variants of \Trainer by replacing our combinatorial contrastive learning loss with NCE, InfoNCE, and CE loss, and follow the same training, validation, and testing dataset in RQ1 for evaluation. To answer RQ3b, we also respectively build three variants of \Explainer by replacing our dual-view causal inference with GNNExplainer, PGExplainer, and CF-GNNExplainer, and adopt the same evaluation dataset in RQ2 for evaluation.

\underline{\emph{Evaluation Metrics.}}
We use the same metrics as in RQ1 and RQ2.

\subsubsection{RQ3a: Impact of Combinatorial Contrastive Learning.}
Table \ref{Detection results}   presents the experimental results of \Trainer (Ours) and its variants trained under different loss functions (Column "Loss"). The results demonstrate the contribution of our combinatorial contrastive learning to the overall detection performance of \Trainer. In particular, we can observe that detection models which are trained with traditional cross-entropy loss outperform their variants trained with self-supervised contrastive loss (InfoNCE and NCE). It is reasonable because fine-tuning on the labeled vulnerability dataset may significantly alter the distribution of learned feature representations. As a result, the robustness and accuracy of learned deep representations enhanced by self-supervised pre-training may not longer hold after supervised fine-tuning. On the contrary, the supervised contrastive learning allows us to effectively leverage label information, which groups the samples belonging to the same class as well as the semantically equivalent variants while simultaneously pushing away the dissimilar samples. Accordingly, the downstream task-specific generalization and robustness can be retained as much as possible.

\begin{table}[t]
 \caption{Contributions of different explanation approaches.}
  \centering
  \renewcommand\arraystretch{0.75}
  \begin{tabular}{c|l|ccc}
    \toprule
    \textbf{Detector} & \textbf{Approach}     & \textbf{MSP}    & \textbf{MSR}    & \textbf{MIoU}\\
    \midrule
    \multirow{4}*{Devign} & GNNExplainer  & 21.40 & 43.28 & 14.68\\
    &PGExplainer       & 25.39 & 47.86 & 20.17  \\
    &CF-GNNExplainer   & 34.10 & 29.65 & 22.79\\
    &\Explainer               & 43.61 & 52.98 & 39.64 \\
    \midrule
    \multirow{4}*{ReVeal} & GNNExplainer   &   23.06 & 47.28 & 17.11\\
    &PGExplainer       & 26.84 & 51.34 & 21.39\\
    &CF-GNNExplainer   & 39.11 & 34.72 & 28.66 \\
    &\Explainer               & 49.52 & 58.39 & 44.97
    \\
    \midrule
    \multirow{4}*{DeepWuKong} & GNNExplainer  & 18.40 & 37.15 & 16.97\\
    &PGExplainer       & 25.56 & 46.81 & 22.64\\
    &CF-GNNExplainer   & 36.79 & 27.09 & 23.96\\
    &\Explainer               & 40.33 & 47.61 & 34.22 \\
    \bottomrule
  \end{tabular}
  \label{RQ3b}
\end{table}

\subsubsection{RQ3b: Impact of Dual-View Causal Inference.}
\begin{figure}[t]
  \centering
  \includegraphics[width=\linewidth]{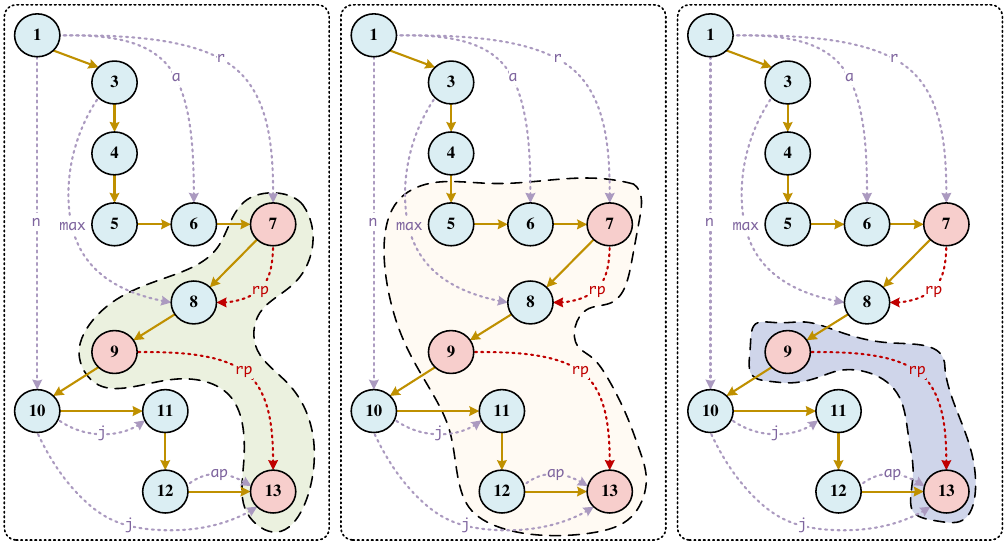}
\caption{\label{ExplanationSample}The PDG sub-graphs (shaded areas) induced by (a) \Explainer, (b) factual reasoning (GNNExplainer), and (c) counterfactual reasoning (CF-GNNExplainer), respectively. Ground truths (i.e., vulnerable nodes and edges) are highlighted in red.}
\end{figure}

Table \ref{RQ3b} shows the performance of \Explainer and its three variants. The results demonstrate that our dual-view causal inference positively contributes to vulnerability explanation. Taking the best performed detection model ReVeal as an example, our \Explainer improves GNNExplainer, PGExplainer, and CF-GNNExplainer by 114.74\%, 84.50\%, and 26.62\% respectively, in terms of MSP, by 23.50\%, 13.73\%, and 68.17\% respectively, in terms of MSR, and by 162.83\%, 110.24\%, and 56.91\% respectively, in terms of MIoU. The results demonstrate the importance of combining factual with counterfactual reasoning for generating both concise and effective explanations. For a more intuitive understanding, we still take the case in qualitative study (Figure \ref{CaseStudy}) as an example. We employ the DeepWuKong, which adopts PDGs as input representations, as the vulnerability detector to derive sub-graph explanations. As shown in Figure \ref{ExplanationSample}, factual-based approach GNNExplainer reveals more rich vulnerability-contexts but also covers redundant nodes and edges, while counterfactual-based approach CF-GNNExplainer has more precise prediction but tends to be conservative and low in coverage. By contrast, \sysname presents all potential vulnerable statements within an accepted scope. In addition, we can find that factual reasoning-based approaches (GNNExplainer and PGExplainer) are higher in MSR, while counterfactual reasoning-based approach (CF-GNNExplainer) is higher in MSP when comparing with each other. This observation further confirms the necessity of combining the strengths of factual and counterfactual reasoning while mitigating each others' weaknesses.

\begin{tcolorbox}[left=1pt,right=1pt,top=1pt,bottom=1pt,boxrule=0.7pt,enhanced,drop fuzzy shadow]
\textbf{Answer to RQ3: }Combinatorial contrastive learning and dual-view causal inference play different roles in our explanation. Combining them together can produce significant improvements.
\end{tcolorbox}

\section{Discussion}
\subsection{Preliminary User Study}
To elaborate the practical value of \sysname, we further perform a small-scale user study to investigate whether \emph{effective} and \emph{concise} explanations can provide more insights and information to help following analysis and repair. Considering a practical application pipeline, we integrated \sysname into DeepWuKong, the best-performing model in RQ1 \& RQ2, to generate explanations.

\underline{\emph{Participants.}}
We invite three MS students with two to five years of experience in developing medium/large-scale C/C++ projects or interning in some security companies for a period of time as our participants. We also invite two security experts from a prominent IT enterprise with at least five years of experience in software security to participate in our user study.

\underline{\emph{Experiment Tasks.}}
We randomly selected 50 vulnerable functions from testing sets in RQ2, and independently assigned 10 unique samples to each participant. For each sample, we present its descriptions and vulnerability-contexts annotated by D2A as well as their corresponding explanations provided by \sysname. The participants are then asked to answer 1) whether the explanation covers enough information to understand the vulnerability; and 2) whether the explanation is concise enough to make further decisions. We use 4-point likert scale \cite{likert1932technique} (1-disagree; 2-somewhat disagree; 3-somewhat agree; 4-agree) to measure the difficulties.

\underline{\emph{Results.}}
Overall, our user study reveals, to some extent, that \sysname presents as many truly vulnerable statements as possible within an accepted scope to help security practitioners understand the detected vulnerability. For \emph{effectiveness} of \sysname, 86\% of the answers are positive (i.e., score $\geq 3$), 12\% are 2 (somewhat disagree), and 2\% are 1 (disagree). For \emph{conciseness} of \sysname, only three (6\%) responses are negative (i.e., score $\leq 2$), which means that explanations provided by \sysname can help them intuitively understand the vulnerable code without checking numerous irrelevant alarms.

\subsection{Threats to Validity}
\textbf{Threats to Internal Validity} come from the quality of our experimental datasets. We evaluate the detection and explanation performance of \sysname on five widely-used real-world benchmarks, and the annotated D2A dataset, respectively. However, existing vulnerability datasets have been reported to exhibit varying degrees of quality issues such as noisy labels and duplication. To reduce the likelihood of experiment biases, following Croft et al.'s \cite{DBLP:conf/icse/CroftBK23} standard practice, we employ two experienced security experts to manually confirm the correctness of vulnerability labels, and leverage a code clone detector to remove duplicate samples.

\noindent \textbf{Threats to External Validity} refer to the generalizability of our approach. We only conduct our experiments on C/C++ datasets, and thus our experimental results may not generalizable to other programming languages such as Java and Python. To mitigate the threat, we employ \code{tree-sitter}, which supports a wide range of languages, to implement \sysname and baselines.

\noindent \textbf{Threats to Construct Validity} refer to the suitability of evaluation measures used for quantifying the performance of vulnerability explanation. We mainly adopt the same metrics following a recent work regarding DL-based vulnerability detectors assessment \cite{BTP}. In the future, we plan to employ other metrics, such as Consistency and Stability \cite{DBLP:conf/issta/HuWLPWZ023,DBLP:conf/eurosp/WarneckeAWR20}, for more comprehensive evaluation.

\section{Related Work}
\subsection{DL-based Vulnerability Detection}
Prior works focus on representing source code as sequences and use LSTM-like models to learn the syntactic and semantic information of vulnerabilities \cite{VulDeePecker,SySeVR,VulDeeLocator,DBLP:journals/kbs/WuSZLSS23}. Recently, a large number of works \cite{Devign,REVEAL,FUNDED,DeepWukong,DBLP:journals/corr/abs-2302-04675,DeepVD} turn to leveraging GNNs to extract rich and well-defined semantics of the program structure from graph representations for downstream vulnerability detection tasks. For example, AMPLE \cite{DBLP:journals/corr/abs-2302-04675} simplifies the input program graph to alleviate the long-term dependency problems and fuses local and global heterogeneous node relations for better representation learning.

In contrast to these studies that aim to design novel neural models for effective vulnerability detection, our goal is to explain their decision logic in a \emph{model-independent} manner. Thus, existing GNN-based approaches are orthogonal to our work and could be adopted together for developing more practical security systems.

\subsection{Explainability on Models of Code}
The requirement for explainability is more urgent in security-related applications \cite{DBLP:journals/tifs/FanWXLGL21,DBLP:conf/eurosp/WarneckeAWR20,DBLP:journals/corr/abs-2208-10605} because it is hard to establish trust on the system decision from simple binary (vulnerable or benign) results without credible evidence. As the most representative attempt, IVDetect \cite{IVDETECT} builds an additional model based on binary detection results to derive crucial statements that are most relevant to the detected vulnerability as explanations. Chakraborty et al. \cite{REVEAL} adopted LEMNA \cite{DBLP:conf/ccs/GuoMXSWX18} to compute the contribution of each code token towards the prediction.


Our approach falls into the category of local explainability, more specifically, perturbation-based approach. A key difference is existing approaches mostly generate explanations from a single view (either factual or counterfactual reasoning) and cannot satisfy special concerns in security domains. By contrast, \sysname proposes dual-view causal inference, which combines the strengths of factual and counterfactual reasoning while mitigating each others’ weaknesses, to provide both effective and concise explanations.


\section{Conclusion and Future Work}
In this paper, we propose \sysname, a general framework to improve and explain GNN-based vulnerability detection systems. Using a combinatorial contrastive learning-based training scheme and a dual-view causal inference-based explanation approach, \sysname is designed to 1) enhance the robustness of existing neural vulnerability detection models to avoid spurious explanations, and 2) provide both concise and effective explanations to reason about the detected vulnerabilities. By applying and evaluating \sysname over three typical GNN-based vulnerability detection models, we show that \sysname can effectively improve the performance of existing GNN-based vulnerability detection models, and provide high-quality explanations.

In the future, we plan to explore a more automated data augmentation approach to further improve the robustness of DL-based detection models. In addition, we aim to work with our industry partners to deploy \sysname in their proprietary security systems to test its effectiveness in practice.

\begin{acks}
This research is supported by the National Natural Science Foundation of China (No.62202414, No.61972335, and No.62002309), the Six Talent Peaks Project in Jiangsu Province (No. RJFW-053); the Jiangsu ``333'' Project and Yangzhou University Top-level Talents Support Program (2019), Postgraduate Research \& Practice Innovation Program of Jiangsu Province (KYCX22\_3502), the Open Funds of State Key Laboratory for Novel Software Technology of Nanjing University (No.KFKT2022B17), the Open Foundation of Yunnan Key Laboratory of Software Engineering (No.2023SE201), the China Scholarship Council Foundation (Nos. 202209300005, 202308320436), and the National Research Foundation, under its Investigatorship Grant (NRF-NRFI08-2022-0002). Any opinions, findings and conclusions or recommendations expressed in this material are those of the author(s) and do not reflect the views of National Research Foundation, Singapore.
\end{acks}

\balance
\bibliographystyle{ACM-Reference-Format}
\bibliography{sample-base}


\begin{thebibliography}{83}


\ifx \showCODEN    \undefined \def \showCODEN     #1{\unskip}     \fi
\ifx \showDOI      \undefined \def \showDOI       #1{#1}\fi
\ifx \showISBNx    \undefined \def \showISBNx     #1{\unskip}     \fi
\ifx \showISBNxiii \undefined \def \showISBNxiii  #1{\unskip}     \fi
\ifx \showISSN     \undefined \def \showISSN      #1{\unskip}     \fi
\ifx \showLCCN     \undefined \def \showLCCN      #1{\unskip}     \fi
\ifx \shownote     \undefined \def \shownote      #1{#1}          \fi
\ifx \showarticletitle \undefined \def \showarticletitle #1{#1}   \fi
\ifx \showURL      \undefined \def \showURL       {\relax}        \fi
\providecommand\bibfield[2]{#2}
\providecommand\bibinfo[2]{#2}
\providecommand\natexlab[1]{#1}
\providecommand\showeprint[2][]{arXiv:#2}

\bibitem[\protect\citeauthoryear{Arcuri and Briand}{Arcuri and Briand}{2011}]%
        {DBLP:conf/icse/ArcuriB11}
\bibfield{author}{\bibinfo{person}{Andrea Arcuri} {and}
  \bibinfo{person}{Lionel~C. Briand}.} \bibinfo{year}{2011}\natexlab{}.
\newblock \showarticletitle{A practical guide for using statistical tests to
  assess randomized algorithms in software engineering}. In
  \bibinfo{booktitle}{\emph{Proceedings of the 33rd International Conference on
  Software Engineering ({ICSE})}}. \bibinfo{publisher}{{ACM}},
  \bibinfo{pages}{1--10}.
\newblock


\bibitem[\protect\citeauthoryear{Bhandari, Naseer, and Moonen}{Bhandari
  et~al\mbox{.}}{2021}]%
        {CVEfixes}
\bibfield{author}{\bibinfo{person}{Guru~Prasad Bhandari},
  \bibinfo{person}{Amara Naseer}, {and} \bibinfo{person}{Leon Moonen}.}
  \bibinfo{year}{2021}\natexlab{}.
\newblock \showarticletitle{CVEfixes: automated collection of vulnerabilities
  and their fixes from open-source software}. In
  \bibinfo{booktitle}{\emph{Proceedings of the 17th International Conference on
  Predictive Models and Data Analytics in Software Engineering (PROMISE)}}.
  \bibinfo{publisher}{{ACM}}, \bibinfo{pages}{30--39}.
\newblock


\bibitem[\protect\citeauthoryear{Bielik and Vechev}{Bielik and Vechev}{2020}]%
        {DBLP:conf/icml/BielikV20}
\bibfield{author}{\bibinfo{person}{Pavol Bielik} {and}
  \bibinfo{person}{Martin~T. Vechev}.} \bibinfo{year}{2020}\natexlab{}.
\newblock \showarticletitle{Adversarial Robustness for Code}. In
  \bibinfo{booktitle}{\emph{Proceedings of the 37th International Conference on
  Machine Learning ({ICML})}}, Vol.~\bibinfo{volume}{119}.
  \bibinfo{pages}{896--907}.
\newblock


\bibitem[\protect\citeauthoryear{Bui, Yu, and Jiang}{Bui et~al\mbox{.}}{2021}]%
        {DBLP:conf/sigir/BuiYJ21}
\bibfield{author}{\bibinfo{person}{Nghi D.~Q. Bui}, \bibinfo{person}{Yijun Yu},
  {and} \bibinfo{person}{Lingxiao Jiang}.} \bibinfo{year}{2021}\natexlab{}.
\newblock \showarticletitle{Self-Supervised Contrastive Learning for Code
  Retrieval and Summarization via Semantic-Preserving Transformations}. In
  \bibinfo{booktitle}{\emph{Proceedings of the 44th International {ACM} {SIGIR}
  Conference on Research and Development in Information Retrieval (SIGIR)}}.
  \bibinfo{publisher}{{ACM}}, \bibinfo{pages}{511--521}.
\newblock


\bibitem[\protect\citeauthoryear{Cao, He, Sun, Ouyang, Zhang, Wu, Su, Bo, Li,
  Ma, Li, and Wei}{Cao et~al\mbox{.}}{2023a}]%
        {DBLP:conf/sp/CaoHSOZWSBLMLW23}
\bibfield{author}{\bibinfo{person}{Sicong Cao}, \bibinfo{person}{Biao He},
  \bibinfo{person}{Xiaobing Sun}, \bibinfo{person}{Yu Ouyang},
  \bibinfo{person}{Chao Zhang}, \bibinfo{person}{Xiaoxue Wu},
  \bibinfo{person}{Ting Su}, \bibinfo{person}{Lili Bo}, \bibinfo{person}{Bin
  Li}, \bibinfo{person}{Chuanlei Ma}, \bibinfo{person}{Jiajia Li}, {and}
  \bibinfo{person}{Tao Wei}.} \bibinfo{year}{2023}\natexlab{a}.
\newblock \showarticletitle{ODDFuzz: Discovering Java Deserialization
  Vulnerabilities via Structure-Aware Directed Greybox Fuzzing}. In
  \bibinfo{booktitle}{\emph{Proceedings of the 44th {IEEE} Symposium on
  Security and Privacy ({SP})}}. \bibinfo{publisher}{{IEEE}},
  \bibinfo{pages}{2726--2743}.
\newblock


\bibitem[\protect\citeauthoryear{Cao, Sun, Bo, Wei, and Li}{Cao
  et~al\mbox{.}}{2021}]%
        {BGNN4VD}
\bibfield{author}{\bibinfo{person}{Sicong Cao}, \bibinfo{person}{Xiaobing Sun},
  \bibinfo{person}{Lili Bo}, \bibinfo{person}{Ying Wei}, {and}
  \bibinfo{person}{Bin Li}.} \bibinfo{year}{2021}\natexlab{}.
\newblock \showarticletitle{\emph{BGNN4VD}: Constructing Bidirectional Graph
  Neural-Network for Vulnerability Detection}.
\newblock \bibinfo{journal}{\emph{Inf. Softw. Technol.}}  \bibinfo{volume}{136}
  (\bibinfo{year}{2021}), \bibinfo{pages}{106576}.
\newblock


\bibitem[\protect\citeauthoryear{Cao, Sun, Bo, Wu, Li, and Tao}{Cao
  et~al\mbox{.}}{2022}]%
        {MVD}
\bibfield{author}{\bibinfo{person}{Sicong Cao}, \bibinfo{person}{Xiaobing Sun},
  \bibinfo{person}{Lili Bo}, \bibinfo{person}{Rongxin Wu}, \bibinfo{person}{Bin
  Li}, {and} \bibinfo{person}{Chuanqi Tao}.} \bibinfo{year}{2022}\natexlab{}.
\newblock \showarticletitle{{MVD:} Memory-Related Vulnerability Detection Based
  on Flow-Sensitive Graph Neural Networks}. In
  \bibinfo{booktitle}{\emph{Proceedings of the 44th {IEEE/ACM} International
  Conference on Software Engineering (ICSE)}}. \bibinfo{publisher}{{ACM}},
  \bibinfo{pages}{1456--1468}.
\newblock


\bibitem[\protect\citeauthoryear{Cao, Sun, Bo, Wu, Li, Wu, Tao, Zhang, and
  Liu}{Cao et~al\mbox{.}}{2024}]%
        {MVD2}
\bibfield{author}{\bibinfo{person}{Sicong Cao}, \bibinfo{person}{Xiaobing Sun},
  \bibinfo{person}{Lili Bo}, \bibinfo{person}{Rongxin Wu}, \bibinfo{person}{Bin
  Li}, \bibinfo{person}{Xiaoxue Wu}, \bibinfo{person}{Chuanqi Tao},
  \bibinfo{person}{Tao Zhang}, {and} \bibinfo{person}{Wei Liu}.}
  \bibinfo{year}{2024}\natexlab{}.
\newblock \showarticletitle{Learning to Detect Memory-related Vulnerabilities}.
\newblock \bibinfo{journal}{\emph{{ACM} Trans. Softw. Eng. Methodol.}}
  \bibinfo{volume}{33}, \bibinfo{number}{2} (\bibinfo{year}{2024}),
  \bibinfo{pages}{43:1--43:35}.
\newblock


\bibitem[\protect\citeauthoryear{Cao, Sun, Wu, Bo, Li, Wu, Liu, He, Ouyang, and
  Li}{Cao et~al\mbox{.}}{2023b}]%
        {DBLP:conf/icse/CaoSWBLWLHOL23}
\bibfield{author}{\bibinfo{person}{Sicong Cao}, \bibinfo{person}{Xiaobing Sun},
  \bibinfo{person}{Xiaoxue Wu}, \bibinfo{person}{Lili Bo}, \bibinfo{person}{Bin
  Li}, \bibinfo{person}{Rongxin Wu}, \bibinfo{person}{Wei Liu},
  \bibinfo{person}{Biao He}, \bibinfo{person}{Yu Ouyang}, {and}
  \bibinfo{person}{Jiajia Li}.} \bibinfo{year}{2023}\natexlab{b}.
\newblock \showarticletitle{Improving Java Deserialization Gadget Chain Mining
  via Overriding-Guided Object Generation}. In
  \bibinfo{booktitle}{\emph{Proceedings of the 45th {IEEE/ACM} International
  Conference on Software Engineering ({ICSE})}}. \bibinfo{publisher}{{IEEE}},
  \bibinfo{pages}{397--409}.
\newblock


\bibitem[\protect\citeauthoryear{Chakraborty, Ahmed, Ding, Devanbu, and
  Ray}{Chakraborty et~al\mbox{.}}{2022a}]%
        {DBLP:conf/sigsoft/ChakrabortyADDR22}
\bibfield{author}{\bibinfo{person}{Saikat Chakraborty},
  \bibinfo{person}{Toufique Ahmed}, \bibinfo{person}{Yangruibo Ding},
  \bibinfo{person}{Premkumar~T. Devanbu}, {and} \bibinfo{person}{Baishakhi
  Ray}.} \bibinfo{year}{2022}\natexlab{a}.
\newblock \showarticletitle{NatGen: generative pre-training by "naturalizing"
  source code}. In \bibinfo{booktitle}{\emph{Proceedings of the 30th {ACM}
  Joint European Software Engineering Conference and Symposium on the
  Foundations of Software Engineering ({ESEC/FSE})}}.
  \bibinfo{publisher}{{ACM}}, \bibinfo{pages}{18--30}.
\newblock


\bibitem[\protect\citeauthoryear{Chakraborty, Krishna, Ding, and
  Ray}{Chakraborty et~al\mbox{.}}{2022b}]%
        {REVEAL}
\bibfield{author}{\bibinfo{person}{Saikat Chakraborty}, \bibinfo{person}{Rahul
  Krishna}, \bibinfo{person}{Yangruibo Ding}, {and} \bibinfo{person}{Baishakhi
  Ray}.} \bibinfo{year}{2022}\natexlab{b}.
\newblock \showarticletitle{Deep Learning based Vulnerability Detection: Are We
  There Yet?}
\newblock \bibinfo{journal}{\emph{{IEEE} Trans. Software Eng.}}
  \bibinfo{volume}{48}, \bibinfo{number}{9} (\bibinfo{year}{2022}),
  \bibinfo{pages}{3280 -- 3296}.
\newblock


\bibitem[\protect\citeauthoryear{Checkmarx}{Checkmarx}{2023}]%
        {Checkmarx}
\bibfield{author}{\bibinfo{person}{Checkmarx}.}
  \bibinfo{year}{2023}\natexlab{}.
\newblock
\newblock
\newblock
\shownote{\url{https://www.checkmarx.com/}.}


\bibitem[\protect\citeauthoryear{Chen, Kornblith, Norouzi, and Hinton}{Chen
  et~al\mbox{.}}{2020}]%
        {NCE}
\bibfield{author}{\bibinfo{person}{Ting Chen}, \bibinfo{person}{Simon
  Kornblith}, \bibinfo{person}{Mohammad Norouzi}, {and}
  \bibinfo{person}{Geoffrey~E. Hinton}.} \bibinfo{year}{2020}\natexlab{}.
\newblock \showarticletitle{A Simple Framework for Contrastive Learning of
  Visual Representations}. In \bibinfo{booktitle}{\emph{Proceedings of the 37th
  International Conference on Machine Learning ({ICML})}},
  Vol.~\bibinfo{volume}{119}. \bibinfo{pages}{1597--1607}.
\newblock


\bibitem[\protect\citeauthoryear{Chen, Hellendoorn, Lamblin, Maniatis,
  Manzagol, Tarlow, and Moitra}{Chen et~al\mbox{.}}{2021}]%
        {DBLP:conf/nips/ChenHLMMTM21}
\bibfield{author}{\bibinfo{person}{Zimin Chen}, \bibinfo{person}{Vincent~J.
  Hellendoorn}, \bibinfo{person}{Pascal Lamblin}, \bibinfo{person}{Petros
  Maniatis}, \bibinfo{person}{Pierre{-}Antoine Manzagol},
  \bibinfo{person}{Daniel Tarlow}, {and} \bibinfo{person}{Subhodeep Moitra}.}
  \bibinfo{year}{2021}\natexlab{}.
\newblock \showarticletitle{{PLUR:} {A} Unifying, Graph-Based View of Program
  Learning, Understanding, and Repair}. In
  \bibinfo{booktitle}{\emph{Proceedings of the 34th Annual Conference on Neural
  Information Processing Systems (NeurIPS)}}. \bibinfo{pages}{23089--23101}.
\newblock


\bibitem[\protect\citeauthoryear{Cheng, Nie, Ningke, Wang, Zheng, and
  Sui}{Cheng et~al\mbox{.}}{2022a}]%
        {BTP}
\bibfield{author}{\bibinfo{person}{Xiao Cheng}, \bibinfo{person}{Xu Nie},
  \bibinfo{person}{Li Ningke}, \bibinfo{person}{Haoyu Wang},
  \bibinfo{person}{Zheng Zheng}, {and} \bibinfo{person}{Yulei Sui}.}
  \bibinfo{year}{2022}\natexlab{a}.
\newblock \showarticletitle{How About Bug-Triggering Paths?-Understanding and
  Characterizing Learning-Based Vulnerability Detectors}.
\newblock \bibinfo{journal}{\emph{{IEEE} Trans. Dependable Secur. Comput.}}
  (\bibinfo{year}{2022}).
\newblock


\bibitem[\protect\citeauthoryear{Cheng, Wang, Hua, Xu, and Sui}{Cheng
  et~al\mbox{.}}{2021}]%
        {DeepWukong}
\bibfield{author}{\bibinfo{person}{Xiao Cheng}, \bibinfo{person}{Haoyu Wang},
  \bibinfo{person}{Jiayi Hua}, \bibinfo{person}{Guoai Xu}, {and}
  \bibinfo{person}{Yulei Sui}.} \bibinfo{year}{2021}\natexlab{}.
\newblock \showarticletitle{DeepWukong: Statically Detecting Software
  Vulnerabilities Using Deep Graph Neural Network}.
\newblock \bibinfo{journal}{\emph{{ACM} Trans. Softw. Eng. Methodol.}}
  \bibinfo{volume}{30}, \bibinfo{number}{3} (\bibinfo{year}{2021}),
  \bibinfo{pages}{38:1--38:33}.
\newblock


\bibitem[\protect\citeauthoryear{Cheng, Zhang, Wang, and Sui}{Cheng
  et~al\mbox{.}}{2022b}]%
        {ContraFlow}
\bibfield{author}{\bibinfo{person}{Xiao Cheng}, \bibinfo{person}{Guanqin
  Zhang}, \bibinfo{person}{Haoyu Wang}, {and} \bibinfo{person}{Yulei Sui}.}
  \bibinfo{year}{2022}\natexlab{b}.
\newblock \showarticletitle{Path-sensitive code embedding via contrastive
  learning for software vulnerability detection}. In
  \bibinfo{booktitle}{\emph{Proceedings of the 31st {ACM} {SIGSOFT}
  International Symposium on Software Testing and Analysis (ISSTA)}}.
  \bibinfo{publisher}{{ACM}}, \bibinfo{pages}{519--531}.
\newblock


\bibitem[\protect\citeauthoryear{Cito, Dillig, Murali, and Chandra}{Cito
  et~al\mbox{.}}{2022}]%
        {DBLP:conf/icse/CitoDMC22}
\bibfield{author}{\bibinfo{person}{J{\"{u}}rgen Cito}, \bibinfo{person}{Isil
  Dillig}, \bibinfo{person}{Vijayaraghavan Murali}, {and}
  \bibinfo{person}{Satish Chandra}.} \bibinfo{year}{2022}\natexlab{}.
\newblock \showarticletitle{Counterfactual Explanations for Models of Code}. In
  \bibinfo{booktitle}{\emph{Proceedings of the 44th {IEEE/ACM} International
  Conference on Software Engineering: Software Engineering in Practice
  ({ICSE}-{SEIP})}}. \bibinfo{publisher}{{IEEE}}, \bibinfo{pages}{125--134}.
\newblock


\bibitem[\protect\citeauthoryear{Croft, Babar, and Kholoosi}{Croft
  et~al\mbox{.}}{2023}]%
        {DBLP:conf/icse/CroftBK23}
\bibfield{author}{\bibinfo{person}{Roland Croft}, \bibinfo{person}{Muhammad~Ali
  Babar}, {and} \bibinfo{person}{M.~Mehdi Kholoosi}.}
  \bibinfo{year}{2023}\natexlab{}.
\newblock \showarticletitle{Data Quality for Software Vulnerability Datasets}.
  In \bibinfo{booktitle}{\emph{Proceedings of the 45th {IEEE/ACM} International
  Conference on Software Engineering (ICSE)}}. \bibinfo{publisher}{{IEEE}},
  \bibinfo{pages}{121--133}.
\newblock


\bibitem[\protect\citeauthoryear{Dam, Tran, and Ghose}{Dam
  et~al\mbox{.}}{2018}]%
        {DBLP:conf/icse/Dam0G18}
\bibfield{author}{\bibinfo{person}{Hoa~Khanh Dam}, \bibinfo{person}{Truyen
  Tran}, {and} \bibinfo{person}{Aditya Ghose}.}
  \bibinfo{year}{2018}\natexlab{}.
\newblock \showarticletitle{Explainable software analytics}. In
  \bibinfo{booktitle}{\emph{Proceedings of the 40th International Conference on
  Software Engineering: New Ideas and Emerging Results ({ICSE}-{NIER})}}.
  \bibinfo{publisher}{{ACM}}, \bibinfo{pages}{53--56}.
\newblock


\bibitem[\protect\citeauthoryear{Dam, Tran, Pham, Ng, Grundy, and Ghose}{Dam
  et~al\mbox{.}}{2021}]%
        {TOKEN}
\bibfield{author}{\bibinfo{person}{Hoa~Khanh Dam}, \bibinfo{person}{Truyen
  Tran}, \bibinfo{person}{Trang Pham}, \bibinfo{person}{Shien~Wee Ng},
  \bibinfo{person}{John Grundy}, {and} \bibinfo{person}{Aditya Ghose}.}
  \bibinfo{year}{2021}\natexlab{}.
\newblock \showarticletitle{Automatic Feature Learning for Predicting
  Vulnerable Software Components}.
\newblock \bibinfo{journal}{\emph{{IEEE} Trans. Software Eng.}}
  \bibinfo{volume}{47}, \bibinfo{number}{1} (\bibinfo{year}{2021}),
  \bibinfo{pages}{67--85}.
\newblock


\bibitem[\protect\citeauthoryear{Ding, Buratti, Pujar, Morari, Ray, and
  Chakraborty}{Ding et~al\mbox{.}}{2022}]%
        {DBLP:conf/acl/DingBPMRC22}
\bibfield{author}{\bibinfo{person}{Yangruibo Ding}, \bibinfo{person}{Luca
  Buratti}, \bibinfo{person}{Saurabh Pujar}, \bibinfo{person}{Alessandro
  Morari}, \bibinfo{person}{Baishakhi Ray}, {and} \bibinfo{person}{Saikat
  Chakraborty}.} \bibinfo{year}{2022}\natexlab{}.
\newblock \showarticletitle{Towards Learning (Dis)-Similarity of Source Code
  from Program Contrasts}. In \bibinfo{booktitle}{\emph{Proceedings of the 60th
  Annual Meeting of the Association for Computational Linguistics ({ACL})}}.
  \bibinfo{publisher}{Association for Computational Linguistics},
  \bibinfo{pages}{6300--6312}.
\newblock


\bibitem[\protect\citeauthoryear{Ding, Chakraborty, Buratti, Pujar, Morari,
  Kaiser, and Ray}{Ding et~al\mbox{.}}{2023}]%
        {DBLP:conf/issta/DingCBPMKR23}
\bibfield{author}{\bibinfo{person}{Yangruibo Ding}, \bibinfo{person}{Saikat
  Chakraborty}, \bibinfo{person}{Luca Buratti}, \bibinfo{person}{Saurabh
  Pujar}, \bibinfo{person}{Alessandro Morari}, \bibinfo{person}{Gail~E.
  Kaiser}, {and} \bibinfo{person}{Baishakhi Ray}.}
  \bibinfo{year}{2023}\natexlab{}.
\newblock \showarticletitle{{CONCORD:} Clone-Aware Contrastive Learning for
  Source Code}. In \bibinfo{booktitle}{\emph{Proceedings of the 32nd {ACM}
  {SIGSOFT} International Symposium on Software Testing and Analysis
  ({ISSTA})}}. \bibinfo{publisher}{{ACM}}, \bibinfo{pages}{26--38}.
\newblock


\bibitem[\protect\citeauthoryear{Fan, Li, Wang, and Nguyen}{Fan
  et~al\mbox{.}}{2020}]%
        {FAN}
\bibfield{author}{\bibinfo{person}{Jiahao Fan}, \bibinfo{person}{Yi Li},
  \bibinfo{person}{Shaohua Wang}, {and} \bibinfo{person}{Tien~N. Nguyen}.}
  \bibinfo{year}{2020}\natexlab{}.
\newblock \showarticletitle{A {C/C++} Code Vulnerability Dataset with Code
  Changes and {CVE} Summaries}. In \bibinfo{booktitle}{\emph{Proceedings of the
  17th International Conference on Mining Software Repositories (MSR)}}.
  \bibinfo{publisher}{{ACM}}, \bibinfo{pages}{508--512}.
\newblock


\bibitem[\protect\citeauthoryear{Fan, Wei, Xie, Liu, Guan, and Liu}{Fan
  et~al\mbox{.}}{2021}]%
        {DBLP:journals/tifs/FanWXLGL21}
\bibfield{author}{\bibinfo{person}{Ming Fan}, \bibinfo{person}{Wenying Wei},
  \bibinfo{person}{Xiaofei Xie}, \bibinfo{person}{Yang Liu},
  \bibinfo{person}{Xiaohong Guan}, {and} \bibinfo{person}{Ting Liu}.}
  \bibinfo{year}{2021}\natexlab{}.
\newblock \showarticletitle{Can We Trust Your Explanations? Sanity Checks for
  Interpreters in Android Malware Analysis}.
\newblock \bibinfo{journal}{\emph{{IEEE} Trans. Inf. Forensics Secur.}}
  \bibinfo{volume}{16} (\bibinfo{year}{2021}), \bibinfo{pages}{838--853}.
\newblock


\bibitem[\protect\citeauthoryear{Flawfinder}{Flawfinder}{2023}]%
        {Flawfinder}
\bibfield{author}{\bibinfo{person}{Flawfinder}.}
  \bibinfo{year}{2023}\natexlab{}.
\newblock
\newblock
\newblock
\shownote{\url{http://www.dwheeler.com/FlawFinder}.}


\bibitem[\protect\citeauthoryear{Fong and Vedaldi}{Fong and Vedaldi}{2017}]%
        {DBLP:conf/iccv/FongV17}
\bibfield{author}{\bibinfo{person}{Ruth~C. Fong} {and} \bibinfo{person}{Andrea
  Vedaldi}.} \bibinfo{year}{2017}\natexlab{}.
\newblock \showarticletitle{Interpretable Explanations of Black Boxes by
  Meaningful Perturbation}. In \bibinfo{booktitle}{\emph{Proceedings of the
  16th {IEEE} International Conference on Computer Vision ({ICCV})}}.
  \bibinfo{publisher}{{IEEE} Computer Society}, \bibinfo{pages}{3449--3457}.
\newblock


\bibitem[\protect\citeauthoryear{Fu and Tantithamthavorn}{Fu and
  Tantithamthavorn}{2022}]%
        {LineVul}
\bibfield{author}{\bibinfo{person}{Michael Fu} {and} \bibinfo{person}{Chakkrit
  Tantithamthavorn}.} \bibinfo{year}{2022}\natexlab{}.
\newblock \showarticletitle{LineVul: {A} Transformer-based Line-Level
  Vulnerability Prediction}. In \bibinfo{booktitle}{\emph{Proceedings of the
  19th {IEEE/ACM} International Conference on Mining Software Repositories
  ({MSR})}}. \bibinfo{publisher}{{IEEE}}, \bibinfo{pages}{608--620}.
\newblock


\bibitem[\protect\citeauthoryear{Ganz, H{\"{a}}rterich, Warnecke, and
  Rieck}{Ganz et~al\mbox{.}}{2021}]%
        {DBLP:conf/ccs/GanzHWR21}
\bibfield{author}{\bibinfo{person}{Tom Ganz}, \bibinfo{person}{Martin
  H{\"{a}}rterich}, \bibinfo{person}{Alexander Warnecke}, {and}
  \bibinfo{person}{Konrad Rieck}.} \bibinfo{year}{2021}\natexlab{}.
\newblock \showarticletitle{Explaining Graph Neural Networks for Vulnerability
  Discovery}. In \bibinfo{booktitle}{\emph{Proceedings of the 14th {ACM}
  Workshop on Artificial Intelligence and Security (AISec@CCS)}}.
  \bibinfo{publisher}{{ACM}}, \bibinfo{pages}{145--156}.
\newblock


\bibitem[\protect\citeauthoryear{Guo, Mu, Xu, Su, Wang, and Xing}{Guo
  et~al\mbox{.}}{2018}]%
        {DBLP:conf/ccs/GuoMXSWX18}
\bibfield{author}{\bibinfo{person}{Wenbo Guo}, \bibinfo{person}{Dongliang Mu},
  \bibinfo{person}{Jun Xu}, \bibinfo{person}{Purui Su}, \bibinfo{person}{Gang
  Wang}, {and} \bibinfo{person}{Xinyu Xing}.} \bibinfo{year}{2018}\natexlab{}.
\newblock \showarticletitle{{LEMNA:} Explaining Deep Learning based Security
  Applications}. In \bibinfo{booktitle}{\emph{Proceedings of the 25th {ACM}
  {SIGSAC} Conference on Computer and Communications Security ({CCS})}}.
  \bibinfo{publisher}{{ACM}}, \bibinfo{pages}{364--379}.
\newblock


\bibitem[\protect\citeauthoryear{He, Ji, and Huang}{He et~al\mbox{.}}{2022}]%
        {DBLP:conf/eurosp/HeJH22}
\bibfield{author}{\bibinfo{person}{Haoyu He}, \bibinfo{person}{Yuede Ji}, {and}
  \bibinfo{person}{H.~Howie Huang}.} \bibinfo{year}{2022}\natexlab{}.
\newblock \showarticletitle{Illuminati: Towards Explaining Graph Neural
  Networks for Cybersecurity Analysis}. In
  \bibinfo{booktitle}{\emph{Proceedings of the 7th {IEEE} European Symposium on
  Security and Privacy (EuroS{\&}P)}}. \bibinfo{publisher}{{IEEE}},
  \bibinfo{pages}{74--89}.
\newblock


\bibitem[\protect\citeauthoryear{Hin, Kan, Chen, and Babar}{Hin
  et~al\mbox{.}}{2022}]%
        {LineVD}
\bibfield{author}{\bibinfo{person}{David Hin}, \bibinfo{person}{Andrey Kan},
  \bibinfo{person}{Huaming Chen}, {and} \bibinfo{person}{Muhammad~Ali Babar}.}
  \bibinfo{year}{2022}\natexlab{}.
\newblock \showarticletitle{LineVD: Statement-level Vulnerability Detection
  using Graph Neural Networks}. In \bibinfo{booktitle}{\emph{Proceedings of the
  19th {IEEE/ACM} International Conference on Mining Software Repositories
  ({MSR})}}. \bibinfo{publisher}{{IEEE}}, \bibinfo{pages}{596--607}.
\newblock


\bibitem[\protect\citeauthoryear{Hu, Wang, Li, Peng, Wu, Zou, and Jin}{Hu
  et~al\mbox{.}}{2023}]%
        {DBLP:conf/issta/HuWLPWZ023}
\bibfield{author}{\bibinfo{person}{Yutao Hu}, \bibinfo{person}{Suyuan Wang},
  \bibinfo{person}{Wenke Li}, \bibinfo{person}{Junru Peng},
  \bibinfo{person}{Yueming Wu}, \bibinfo{person}{Deqing Zou}, {and}
  \bibinfo{person}{Hai Jin}.} \bibinfo{year}{2023}\natexlab{}.
\newblock \showarticletitle{Interpreters for GNN-Based Vulnerability Detection:
  Are We There Yet?}. In \bibinfo{booktitle}{\emph{Proceedings of the 32nd
  {ACM} {SIGSOFT} International Symposium on Software Testing and Analysis
  ({ISSTA})}}. \bibinfo{publisher}{{ACM}}, \bibinfo{pages}{1407--1419}.
\newblock


\bibitem[\protect\citeauthoryear{Jain, Jain, Zhang, Abbeel, Gonzalez, and
  Stoica}{Jain et~al\mbox{.}}{2021}]%
        {ContraCode}
\bibfield{author}{\bibinfo{person}{Paras Jain}, \bibinfo{person}{Ajay Jain},
  \bibinfo{person}{Tianjun Zhang}, \bibinfo{person}{Pieter Abbeel},
  \bibinfo{person}{Joseph Gonzalez}, {and} \bibinfo{person}{Ion Stoica}.}
  \bibinfo{year}{2021}\natexlab{}.
\newblock \showarticletitle{Contrastive Code Representation Learning}. In
  \bibinfo{booktitle}{\emph{Proceedings of the 26th Conference on Empirical
  Methods in Natural Language Processing ({EMNLP})}}.
  \bibinfo{publisher}{Association for Computational Linguistics},
  \bibinfo{pages}{5954--5971}.
\newblock


\bibitem[\protect\citeauthoryear{Jia, Srikant, Mitrovska, Gan, Chang, Liu, and
  O'Reilly}{Jia et~al\mbox{.}}{2023}]%
        {DBLP:journals/corr/abs-2211-11711}
\bibfield{author}{\bibinfo{person}{Jinghan Jia}, \bibinfo{person}{Shashank
  Srikant}, \bibinfo{person}{Tamara Mitrovska}, \bibinfo{person}{Chuang Gan},
  \bibinfo{person}{Shiyu Chang}, \bibinfo{person}{Sijia Liu}, {and}
  \bibinfo{person}{Una{-}May O'Reilly}.} \bibinfo{year}{2023}\natexlab{}.
\newblock \showarticletitle{{CLAWSAT:} Towards Both Robust and Accurate Code
  Models}. In \bibinfo{booktitle}{\emph{Proceedings of the 30th {IEEE}
  International Conference on Software Analysis, Evolution and Reengineering
  ({SANER})}}. \bibinfo{publisher}{{IEEE}}, \bibinfo{pages}{212--223}.
\newblock


\bibitem[\protect\citeauthoryear{Johnson, Dempsey, Ross, Gupta, Bailey,
  et~al\mbox{.}}{Johnson et~al\mbox{.}}{2011}]%
        {johnson2011guide}
\bibfield{author}{\bibinfo{person}{Arnold Johnson}, \bibinfo{person}{Kelley
  Dempsey}, \bibinfo{person}{Ron Ross}, \bibinfo{person}{Sarbari Gupta},
  \bibinfo{person}{Dennis Bailey}, {et~al\mbox{.}}}
  \bibinfo{year}{2011}\natexlab{}.
\newblock \showarticletitle{Guide for security-focused configuration management
  of information systems}.
\newblock \bibinfo{journal}{\emph{NIST special publication}}
  \bibinfo{volume}{800}, \bibinfo{number}{128} (\bibinfo{year}{2011}),
  \bibinfo{pages}{16--16}.
\newblock


\bibitem[\protect\citeauthoryear{Khosla, Teterwak, Wang, Sarna, Tian, Isola,
  Maschinot, Liu, and Krishnan}{Khosla et~al\mbox{.}}{2020}]%
        {SupCon}
\bibfield{author}{\bibinfo{person}{Prannay Khosla}, \bibinfo{person}{Piotr
  Teterwak}, \bibinfo{person}{Chen Wang}, \bibinfo{person}{Aaron Sarna},
  \bibinfo{person}{Yonglong Tian}, \bibinfo{person}{Phillip Isola},
  \bibinfo{person}{Aaron Maschinot}, \bibinfo{person}{Ce Liu}, {and}
  \bibinfo{person}{Dilip Krishnan}.} \bibinfo{year}{2020}\natexlab{}.
\newblock \showarticletitle{Supervised Contrastive Learning}. In
  \bibinfo{booktitle}{\emph{Proceedings of the 34th Annual Conference on Neural
  Information Processing Systems (NeurIPS)}}.
\newblock


\bibitem[\protect\citeauthoryear{Kingma and Ba}{Kingma and Ba}{2015}]%
        {ADAM}
\bibfield{author}{\bibinfo{person}{Diederik~P. Kingma} {and}
  \bibinfo{person}{Jimmy Ba}.} \bibinfo{year}{2015}\natexlab{}.
\newblock \showarticletitle{Adam: {A} Method for Stochastic Optimization}. In
  \bibinfo{booktitle}{\emph{Proceedings of the 3rd International Conference on
  Learning Representations (ICLR)}}.
\newblock


\bibitem[\protect\citeauthoryear{Li, Tarlow, Brockschmidt, and Zemel}{Li
  et~al\mbox{.}}{2016}]%
        {GGNN}
\bibfield{author}{\bibinfo{person}{Yujia Li}, \bibinfo{person}{Daniel Tarlow},
  \bibinfo{person}{Marc Brockschmidt}, {and} \bibinfo{person}{Richard~S.
  Zemel}.} \bibinfo{year}{2016}\natexlab{}.
\newblock \showarticletitle{Gated Graph Sequence Neural Networks}. In
  \bibinfo{booktitle}{\emph{Proceedings of the 4th International Conference on
  Learning Representations (ICLR)}}.
\newblock


\bibitem[\protect\citeauthoryear{Li, Wang, and Nguyen}{Li
  et~al\mbox{.}}{2021}]%
        {IVDETECT}
\bibfield{author}{\bibinfo{person}{Yi Li}, \bibinfo{person}{Shaohua Wang},
  {and} \bibinfo{person}{Tien~N. Nguyen}.} \bibinfo{year}{2021}\natexlab{}.
\newblock \showarticletitle{Vulnerability detection with fine-grained
  interpretations}. In \bibinfo{booktitle}{\emph{Proceeding of the 29th {ACM}
  Joint European Software Engineering Conference and Symposium on the
  Foundations of Software Engineering (ESEC/FSE)}}. \bibinfo{publisher}{{ACM}},
  \bibinfo{pages}{292--303}.
\newblock


\bibitem[\protect\citeauthoryear{Li, Zou, Xu, Chen, Zhu, and Jin}{Li
  et~al\mbox{.}}{2022a}]%
        {VulDeeLocator}
\bibfield{author}{\bibinfo{person}{Zhen Li}, \bibinfo{person}{Deqing Zou},
  \bibinfo{person}{Shouhuai Xu}, \bibinfo{person}{Zhaoxuan Chen},
  \bibinfo{person}{Yawei Zhu}, {and} \bibinfo{person}{Hai Jin}.}
  \bibinfo{year}{2022}\natexlab{a}.
\newblock \showarticletitle{VulDeeLocator: {A} Deep Learning-Based Fine-Grained
  Vulnerability Detector}.
\newblock \bibinfo{journal}{\emph{{IEEE} Trans. Dependable Secur. Comput.}}
  \bibinfo{volume}{19}, \bibinfo{number}{4} (\bibinfo{year}{2022}),
  \bibinfo{pages}{2821--2837}.
\newblock


\bibitem[\protect\citeauthoryear{Li, Zou, Xu, Jin, Zhu, and Chen}{Li
  et~al\mbox{.}}{2022b}]%
        {SySeVR}
\bibfield{author}{\bibinfo{person}{Zhen Li}, \bibinfo{person}{Deqing Zou},
  \bibinfo{person}{Shouhuai Xu}, \bibinfo{person}{Hai Jin},
  \bibinfo{person}{Yawei Zhu}, {and} \bibinfo{person}{Zhaoxuan Chen}.}
  \bibinfo{year}{2022}\natexlab{b}.
\newblock \showarticletitle{SySeVR: {A} Framework for Using Deep Learning to
  Detect Software Vulnerabilities}.
\newblock \bibinfo{journal}{\emph{{IEEE} Trans. Dependable Secur. Comput.}}
  \bibinfo{volume}{19}, \bibinfo{number}{4} (\bibinfo{year}{2022}),
  \bibinfo{pages}{2244--2258}.
\newblock


\bibitem[\protect\citeauthoryear{Li, Zou, Xu, Ou, Jin, Wang, Deng, and
  Zhong}{Li et~al\mbox{.}}{2018}]%
        {VulDeePecker}
\bibfield{author}{\bibinfo{person}{Zhen Li}, \bibinfo{person}{Deqing Zou},
  \bibinfo{person}{Shouhuai Xu}, \bibinfo{person}{Xinyu Ou},
  \bibinfo{person}{Hai Jin}, \bibinfo{person}{Sujuan Wang},
  \bibinfo{person}{Zhijun Deng}, {and} \bibinfo{person}{Yuyi Zhong}.}
  \bibinfo{year}{2018}\natexlab{}.
\newblock \showarticletitle{{VulDeePecker}: {A} Deep Learning-Based System for
  Vulnerability Detection}. In \bibinfo{booktitle}{\emph{Proceedings of the
  25th Annual Network and Distributed System Security Symposium (NDSS)}}.
  \bibinfo{publisher}{The Internet Society}.
\newblock


\bibitem[\protect\citeauthoryear{Likert}{Likert}{1932}]%
        {likert1932technique}
\bibfield{author}{\bibinfo{person}{Rensis Likert}.}
  \bibinfo{year}{1932}\natexlab{}.
\newblock \showarticletitle{A technique for the measurement of attitudes.}
\newblock \bibinfo{journal}{\emph{Archives of psychology}}
  (\bibinfo{year}{1932}).
\newblock


\bibitem[\protect\citeauthoryear{Lin, Lan, and Li}{Lin et~al\mbox{.}}{2021}]%
        {DBLP:conf/icml/LinLL21}
\bibfield{author}{\bibinfo{person}{Wanyu Lin}, \bibinfo{person}{Hao Lan}, {and}
  \bibinfo{person}{Baochun Li}.} \bibinfo{year}{2021}\natexlab{}.
\newblock \showarticletitle{Generative Causal Explanations for Graph Neural
  Networks}. In \bibinfo{booktitle}{\emph{Proceedings of the 38th International
  Conference on Machine Learning ({ICML})}}, Vol.~\bibinfo{volume}{139}.
  \bibinfo{pages}{6666--6679}.
\newblock


\bibitem[\protect\citeauthoryear{Liu, Wu, Xie, Meng, and Liu}{Liu
  et~al\mbox{.}}{2023}]%
        {ContraBERT}
\bibfield{author}{\bibinfo{person}{Shangqing Liu}, \bibinfo{person}{Bozhi Wu},
  \bibinfo{person}{Xiaofei Xie}, \bibinfo{person}{Guozhu Meng}, {and}
  \bibinfo{person}{Yang Liu}.} \bibinfo{year}{2023}\natexlab{}.
\newblock \showarticletitle{ContraBERT: Enhancing Code Pre-trained Models via
  Contrastive Learning}. In \bibinfo{booktitle}{\emph{Proceedings of the 45th
  {IEEE/ACM} International Conference on Software Engineering ({ICSE})}}.
  \bibinfo{publisher}{{IEEE}}.
\newblock


\bibitem[\protect\citeauthoryear{Lucic, ter Hoeve, Tolomei, de~Rijke, and
  Silvestri}{Lucic et~al\mbox{.}}{2022}]%
        {CF-GNNExplainer}
\bibfield{author}{\bibinfo{person}{Ana Lucic}, \bibinfo{person}{Maartje~A. ter
  Hoeve}, \bibinfo{person}{Gabriele Tolomei}, \bibinfo{person}{Maarten de
  Rijke}, {and} \bibinfo{person}{Fabrizio Silvestri}.}
  \bibinfo{year}{2022}\natexlab{}.
\newblock \showarticletitle{CF-GNNExplainer: Counterfactual Explanations for
  Graph Neural Networks}. In \bibinfo{booktitle}{\emph{Proceedings of the 25th
  International Conference on Artificial Intelligence and Statistics
  ({AISTATS})}}, Vol.~\bibinfo{volume}{151}. \bibinfo{pages}{4499--4511}.
\newblock


\bibitem[\protect\citeauthoryear{Lundberg and Lee}{Lundberg and Lee}{2017}]%
        {DBLP:conf/nips/LundbergL17}
\bibfield{author}{\bibinfo{person}{Scott~M. Lundberg} {and}
  \bibinfo{person}{Su{-}In Lee}.} \bibinfo{year}{2017}\natexlab{}.
\newblock \showarticletitle{A Unified Approach to Interpreting Model
  Predictions}. In \bibinfo{booktitle}{\emph{Proceedings of the 31st Annual
  Conference on Neural Information Processing Systems (NeurIPS)}}.
  \bibinfo{pages}{4765--4774}.
\newblock


\bibitem[\protect\citeauthoryear{Luo, Cheng, Xu, Yu, Zong, Chen, and Zhang}{Luo
  et~al\mbox{.}}{2020}]%
        {PGExplainer}
\bibfield{author}{\bibinfo{person}{Dongsheng Luo}, \bibinfo{person}{Wei Cheng},
  \bibinfo{person}{Dongkuan Xu}, \bibinfo{person}{Wenchao Yu},
  \bibinfo{person}{Bo Zong}, \bibinfo{person}{Haifeng Chen}, {and}
  \bibinfo{person}{Xiang Zhang}.} \bibinfo{year}{2020}\natexlab{}.
\newblock \showarticletitle{Parameterized Explainer for Graph Neural Network}.
  In \bibinfo{booktitle}{\emph{Proceedings of the 34th Annual Conference on
  Neural Information Processing Systems (NeurIPS)}}.
\newblock


\bibitem[\protect\citeauthoryear{Nadeem, Vos, Cao, Pajola, Dieck, Baumgartner,
  and Verwer}{Nadeem et~al\mbox{.}}{2022}]%
        {DBLP:journals/corr/abs-2208-10605}
\bibfield{author}{\bibinfo{person}{Azqa Nadeem}, \bibinfo{person}{Dani{\"{e}}l
  Vos}, \bibinfo{person}{Clinton Cao}, \bibinfo{person}{Luca Pajola},
  \bibinfo{person}{Simon Dieck}, \bibinfo{person}{Robert Baumgartner}, {and}
  \bibinfo{person}{Sicco Verwer}.} \bibinfo{year}{2022}\natexlab{}.
\newblock \showarticletitle{SoK: Explainable Machine Learning for Computer
  Security Applications}.
\newblock \bibinfo{journal}{\emph{arXiv preprint arXiv:2208.10605}}
  (\bibinfo{year}{2022}).
\newblock


\bibitem[\protect\citeauthoryear{Nikitopoulos, Dritsa, Louridas, and
  Mitropoulos}{Nikitopoulos et~al\mbox{.}}{2021}]%
        {CrossVul}
\bibfield{author}{\bibinfo{person}{Georgios Nikitopoulos},
  \bibinfo{person}{Konstantina Dritsa}, \bibinfo{person}{Panos Louridas}, {and}
  \bibinfo{person}{Dimitris Mitropoulos}.} \bibinfo{year}{2021}\natexlab{}.
\newblock \showarticletitle{CrossVul: a cross-language vulnerability dataset
  with commit data}. In \bibinfo{booktitle}{\emph{Proceedings of the 29th {ACM}
  Joint European Software Engineering Conference and Symposium on the
  Foundations of Software Engineering ({ESEC/FSE})}}.
  \bibinfo{publisher}{{ACM}}, \bibinfo{pages}{1565--1569}.
\newblock


\bibitem[\protect\citeauthoryear{Nong, Ou, Pradel, Chen, and Cai}{Nong
  et~al\mbox{.}}{2022}]%
        {DBLP:conf/sigsoft/NongOP0C22}
\bibfield{author}{\bibinfo{person}{Yu Nong}, \bibinfo{person}{Yuzhe Ou},
  \bibinfo{person}{Michael Pradel}, \bibinfo{person}{Feng Chen}, {and}
  \bibinfo{person}{Haipeng Cai}.} \bibinfo{year}{2022}\natexlab{}.
\newblock \showarticletitle{Generating realistic vulnerabilities via neural
  code editing: an empirical study}. In \bibinfo{booktitle}{\emph{Proceedings
  of the 30th {ACM} Joint European Software Engineering Conference and
  Symposium on the Foundations of Software Engineering ({ESEC/FSE})}}.
  \bibinfo{publisher}{{ACM}}, \bibinfo{pages}{1097--1109}.
\newblock


\bibitem[\protect\citeauthoryear{Nong, Ou, Pradel, Chen, and Cai}{Nong
  et~al\mbox{.}}{2023}]%
        {DBLP:conf/icse/NongOPCC23}
\bibfield{author}{\bibinfo{person}{Yu Nong}, \bibinfo{person}{Yuzhe Ou},
  \bibinfo{person}{Michael Pradel}, \bibinfo{person}{Feng Chen}, {and}
  \bibinfo{person}{Haipeng Cai}.} \bibinfo{year}{2023}\natexlab{}.
\newblock \showarticletitle{{VULGEN:} Realistic Vulnerability Generation Via
  Pattern Mining and Deep Learning}. In \bibinfo{booktitle}{\emph{Proceedings
  of 45th {IEEE/ACM} International Conference on Software Engineering
  ({ICSE})}}. \bibinfo{publisher}{{IEEE}}, \bibinfo{pages}{2527--2539}.
\newblock


\bibitem[\protect\citeauthoryear{Pendleton, Garcia{-}Lebron, Cho, and
  Xu}{Pendleton et~al\mbox{.}}{2017}]%
        {DBLP:journals/csur/PendletonGCX17}
\bibfield{author}{\bibinfo{person}{Marcus Pendleton}, \bibinfo{person}{Richard
  Garcia{-}Lebron}, \bibinfo{person}{Jin{-}Hee Cho}, {and}
  \bibinfo{person}{Shouhuai Xu}.} \bibinfo{year}{2017}\natexlab{}.
\newblock \showarticletitle{A Survey on Systems Security Metrics}.
\newblock \bibinfo{journal}{\emph{{ACM} Comput. Surv.}} \bibinfo{volume}{49},
  \bibinfo{number}{4} (\bibinfo{year}{2017}), \bibinfo{pages}{62:1--62:35}.
\newblock


\bibitem[\protect\citeauthoryear{Pornprasit, Tantithamthavorn, Jiarpakdee, Fu,
  and Thongtanunam}{Pornprasit et~al\mbox{.}}{2021}]%
        {Pyexplainer}
\bibfield{author}{\bibinfo{person}{Chanathip Pornprasit},
  \bibinfo{person}{Chakkrit Tantithamthavorn}, \bibinfo{person}{Jirayus
  Jiarpakdee}, \bibinfo{person}{Michael Fu}, {and} \bibinfo{person}{Patanamon
  Thongtanunam}.} \bibinfo{year}{2021}\natexlab{}.
\newblock \showarticletitle{PyExplainer: Explaining the Predictions of
  Just-In-Time Defect Models}. In \bibinfo{booktitle}{\emph{Proceedings of the
  36th {IEEE/ACM} International Conference on Automated Software Engineering
  ({ASE})}}. \bibinfo{publisher}{{IEEE}}, \bibinfo{pages}{407--418}.
\newblock


\bibitem[\protect\citeauthoryear{Rabin, Hellendoorn, and Alipour}{Rabin
  et~al\mbox{.}}{2021}]%
        {DBLP:conf/sigsoft/RabinHA21}
\bibfield{author}{\bibinfo{person}{Md. Rafiqul~Islam Rabin},
  \bibinfo{person}{Vincent~J. Hellendoorn}, {and}
  \bibinfo{person}{Mohammad~Amin Alipour}.} \bibinfo{year}{2021}\natexlab{}.
\newblock \showarticletitle{Understanding neural code intelligence through
  program simplification}. In \bibinfo{booktitle}{\emph{Proceedings of the 29th
  {ACM} Joint European Software Engineering Conference and Symposium on the
  Foundations of Software Engineering ({ESEC/FSE})}}.
  \bibinfo{publisher}{{ACM}}, \bibinfo{pages}{441--452}.
\newblock


\bibitem[\protect\citeauthoryear{Ribeiro, Singh, and Guestrin}{Ribeiro
  et~al\mbox{.}}{2016}]%
        {LIME}
\bibfield{author}{\bibinfo{person}{Marco~T{\'{u}}lio Ribeiro},
  \bibinfo{person}{Sameer Singh}, {and} \bibinfo{person}{Carlos Guestrin}.}
  \bibinfo{year}{2016}\natexlab{}.
\newblock \showarticletitle{"Why Should {I} Trust You?": Explaining the
  Predictions of Any Classifier}. In \bibinfo{booktitle}{\emph{Proceedings of
  the 22nd {ACM} {SIGKDD} International Conference on Knowledge Discovery and
  Data Mining ({KDD})}}. \bibinfo{publisher}{{ACM}},
  \bibinfo{pages}{1135--1144}.
\newblock


\bibitem[\protect\citeauthoryear{Shi, Yin, Wang, Lo, Zhang, Xia, Zhao, and
  Xu}{Shi et~al\mbox{.}}{2022}]%
        {DBLP:conf/sigsoft/ShiYW0Z0ZX22}
\bibfield{author}{\bibinfo{person}{Yucen Shi}, \bibinfo{person}{Ying Yin},
  \bibinfo{person}{Zhengkui Wang}, \bibinfo{person}{David Lo},
  \bibinfo{person}{Tao Zhang}, \bibinfo{person}{Xin Xia},
  \bibinfo{person}{Yuhai Zhao}, {and} \bibinfo{person}{Bowen Xu}.}
  \bibinfo{year}{2022}\natexlab{}.
\newblock \showarticletitle{How to better utilize code graphs in semantic code
  search?}. In \bibinfo{booktitle}{\emph{Proceeding of the 30th {ACM} Joint
  European Software Engineering Conference and Symposium on the Foundations of
  Software Engineering (ESEC/FSE)}}. \bibinfo{publisher}{{ACM}},
  \bibinfo{pages}{722--733}.
\newblock


\bibitem[\protect\citeauthoryear{Steenhoek, Rahman, Jiles, and Le}{Steenhoek
  et~al\mbox{.}}{2018}]%
        {Empirical1}
\bibfield{author}{\bibinfo{person}{Benjamin Steenhoek},
  \bibinfo{person}{Md~Mahbubur Rahman}, \bibinfo{person}{Richard Jiles}, {and}
  \bibinfo{person}{Wei Le}.} \bibinfo{year}{2018}\natexlab{}.
\newblock \showarticletitle{An Empirical Study of Deep Learning Models for
  Vulnerability Detection}. In \bibinfo{booktitle}{\emph{Proceedings of the
  45th International Conference on Software Engineering ({ICSE})}}.
  \bibinfo{publisher}{{IEEE/ACM}}.
\newblock


\bibitem[\protect\citeauthoryear{Sun, Ye, Bo, Wu, Wei, Zhang, and Li}{Sun
  et~al\mbox{.}}{2023}]%
        {DBLP:journals/jss/SunYBWWZL23}
\bibfield{author}{\bibinfo{person}{Xiaobing Sun}, \bibinfo{person}{Zhenlei Ye},
  \bibinfo{person}{Lili Bo}, \bibinfo{person}{Xiaoxue Wu},
  \bibinfo{person}{Ying Wei}, \bibinfo{person}{Tao Zhang}, {and}
  \bibinfo{person}{Bin Li}.} \bibinfo{year}{2023}\natexlab{}.
\newblock \showarticletitle{Automatic software vulnerability assessment by
  extracting vulnerability elements}.
\newblock \bibinfo{journal}{\emph{J. Syst. Softw.}}  \bibinfo{volume}{204}
  (\bibinfo{year}{2023}), \bibinfo{pages}{111790}.
\newblock


\bibitem[\protect\citeauthoryear{Suneja, Zheng, Zhuang, Laredo, and
  Morari}{Suneja et~al\mbox{.}}{2021}]%
        {DBLP:conf/sigsoft/SunejaZZLM21}
\bibfield{author}{\bibinfo{person}{Sahil Suneja}, \bibinfo{person}{Yunhui
  Zheng}, \bibinfo{person}{Yufan Zhuang}, \bibinfo{person}{Jim~Alain Laredo},
  {and} \bibinfo{person}{Alessandro Morari}.} \bibinfo{year}{2021}\natexlab{}.
\newblock \showarticletitle{Probing Model Signal-Awareness via
  Prediction-Preserving Input Minimization}. In
  \bibinfo{booktitle}{\emph{Proceedings of the 29th {ACM} Joint European
  Software Engineering Conference and Symposium on the Foundations of Software
  Engineering ({ESEC/FSE})}}. \bibinfo{publisher}{{ACM}},
  \bibinfo{pages}{945--955}.
\newblock


\bibitem[\protect\citeauthoryear{Suter}{Suter}{1990}]%
        {MLP}
\bibfield{author}{\bibinfo{person}{Bruce~W Suter}.}
  \bibinfo{year}{1990}\natexlab{}.
\newblock \showarticletitle{The multilayer perceptron as an approximation to a
  Bayes optimal discriminant function}.
\newblock \bibinfo{journal}{\emph{IEEE transactions on neural networks}}
  \bibinfo{volume}{1}, \bibinfo{number}{4} (\bibinfo{year}{1990}),
  \bibinfo{pages}{291}.
\newblock


\bibitem[\protect\citeauthoryear{Tan, Geng, Fu, Ge, Xu, Li, and Zhang}{Tan
  et~al\mbox{.}}{2022}]%
        {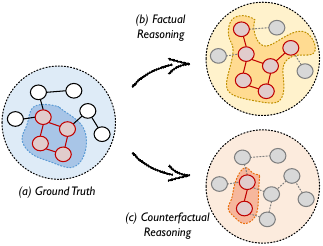}
\bibfield{author}{\bibinfo{person}{Juntao Tan}, \bibinfo{person}{Shijie Geng},
  \bibinfo{person}{Zuohui Fu}, \bibinfo{person}{Yingqiang Ge},
  \bibinfo{person}{Shuyuan Xu}, \bibinfo{person}{Yunqi Li}, {and}
  \bibinfo{person}{Yongfeng Zhang}.} \bibinfo{year}{2022}\natexlab{}.
\newblock \showarticletitle{Learning and Evaluating Graph Neural Network
  Explanations based on Counterfactual and Factual Reasoning}. In
  \bibinfo{booktitle}{\emph{Proceedings of the 31st {ACM} Web Conference
  (WWW)}}. \bibinfo{publisher}{{ACM}}, \bibinfo{pages}{1018--1027}.
\newblock


\bibitem[\protect\citeauthoryear{Tantithamthavorn, McIntosh, Hassan, and
  Matsumoto}{Tantithamthavorn et~al\mbox{.}}{2017}]%
        {DBLP:journals/tse/Tantithamthavorn17}
\bibfield{author}{\bibinfo{person}{Chakkrit Tantithamthavorn},
  \bibinfo{person}{Shane McIntosh}, \bibinfo{person}{Ahmed~E. Hassan}, {and}
  \bibinfo{person}{Kenichi Matsumoto}.} \bibinfo{year}{2017}\natexlab{}.
\newblock \showarticletitle{An Empirical Comparison of Model Validation
  Techniques for Defect Prediction Models}.
\newblock \bibinfo{journal}{\emph{{IEEE} Trans. Software Eng.}}
  \bibinfo{volume}{43}, \bibinfo{number}{1} (\bibinfo{year}{2017}),
  \bibinfo{pages}{1--18}.
\newblock


\bibitem[\protect\citeauthoryear{van~den Oord, Li, and Vinyals}{van~den Oord
  et~al\mbox{.}}{2018}]%
        {InfoNCE}
\bibfield{author}{\bibinfo{person}{A{\"{a}}ron van~den Oord},
  \bibinfo{person}{Yazhe Li}, {and} \bibinfo{person}{Oriol Vinyals}.}
  \bibinfo{year}{2018}\natexlab{}.
\newblock \showarticletitle{Representation Learning with Contrastive Predictive
  Coding}.
\newblock \bibinfo{journal}{\emph{arXiv preprint arXiv:1807.03748}}
  (\bibinfo{year}{2018}).
\newblock


\bibitem[\protect\citeauthoryear{Vaswani, Shazeer, Parmar, Uszkoreit, Jones,
  Gomez, Kaiser, and Polosukhin}{Vaswani et~al\mbox{.}}{2017}]%
        {DBLP:conf/nips/VaswaniSPUJGKP17}
\bibfield{author}{\bibinfo{person}{Ashish Vaswani}, \bibinfo{person}{Noam
  Shazeer}, \bibinfo{person}{Niki Parmar}, \bibinfo{person}{Jakob Uszkoreit},
  \bibinfo{person}{Llion Jones}, \bibinfo{person}{Aidan~N. Gomez},
  \bibinfo{person}{Lukasz Kaiser}, {and} \bibinfo{person}{Illia Polosukhin}.}
  \bibinfo{year}{2017}\natexlab{}.
\newblock \showarticletitle{Attention is All you Need}. In
  \bibinfo{booktitle}{\emph{Proceedings of the 31st Annual Conference on Neural
  Information Processing Systems (NeurIPS)}}. \bibinfo{pages}{5998--6008}.
\newblock


\bibitem[\protect\citeauthoryear{Wang, Ye, Tang, Tan, Huang, Fang, Feng, Bian,
  and Wang}{Wang et~al\mbox{.}}{2021}]%
        {FUNDED}
\bibfield{author}{\bibinfo{person}{Huanting Wang}, \bibinfo{person}{Guixin Ye},
  \bibinfo{person}{Zhanyong Tang}, \bibinfo{person}{Shin~Hwei Tan},
  \bibinfo{person}{Songfang Huang}, \bibinfo{person}{Dingyi Fang},
  \bibinfo{person}{Yansong Feng}, \bibinfo{person}{Lizhong Bian}, {and}
  \bibinfo{person}{Zheng Wang}.} \bibinfo{year}{2021}\natexlab{}.
\newblock \showarticletitle{Combining Graph-Based Learning With Automated Data
  Collection for Code Vulnerability Detection}.
\newblock \bibinfo{journal}{\emph{{IEEE} Trans. Inf. Forensics Secur.}}
  \bibinfo{volume}{16} (\bibinfo{year}{2021}), \bibinfo{pages}{1943--1958}.
\newblock


\bibitem[\protect\citeauthoryear{Wang, Nguyen, Wang, Li, Zhang, and
  Yadavally}{Wang et~al\mbox{.}}{2023}]%
        {DeepVD}
\bibfield{author}{\bibinfo{person}{Wenbo Wang}, \bibinfo{person}{Tien~N.
  Nguyen}, \bibinfo{person}{Shaohua Wang}, \bibinfo{person}{Yi Li},
  \bibinfo{person}{Jiyuan Zhang}, {and} \bibinfo{person}{Aashish Yadavally}.}
  \bibinfo{year}{2023}\natexlab{}.
\newblock \showarticletitle{DeepVD: Toward Class-Separation Features for Neural
  Network Vulnerability Detection}. In \bibinfo{booktitle}{\emph{Proceedings of
  the 45th {IEEE/ACM} International Conference on Software Engineering
  ({ICSE})}}. \bibinfo{publisher}{{IEEE}}.
\newblock


\bibitem[\protect\citeauthoryear{Warnecke, Arp, Wressnegger, and
  Rieck}{Warnecke et~al\mbox{.}}{2020}]%
        {DBLP:conf/eurosp/WarneckeAWR20}
\bibfield{author}{\bibinfo{person}{Alexander Warnecke}, \bibinfo{person}{Daniel
  Arp}, \bibinfo{person}{Christian Wressnegger}, {and} \bibinfo{person}{Konrad
  Rieck}.} \bibinfo{year}{2020}\natexlab{}.
\newblock \showarticletitle{Evaluating Explanation Methods for Deep Learning in
  Security}. In \bibinfo{booktitle}{\emph{Proceedings of the 5th {IEEE}
  European Symposium on Security and Privacy (EuroS{\&}P)}}.
  \bibinfo{publisher}{{IEEE}}, \bibinfo{pages}{158--174}.
\newblock


\bibitem[\protect\citeauthoryear{Wei, Bo, Sun, Li, Zhang, and Tao}{Wei
  et~al\mbox{.}}{2023}]%
        {DBLP:journals/infsof/WeiBSLZT23}
\bibfield{author}{\bibinfo{person}{Ying Wei}, \bibinfo{person}{Lili Bo},
  \bibinfo{person}{Xiaobing Sun}, \bibinfo{person}{Bin Li},
  \bibinfo{person}{Tao Zhang}, {and} \bibinfo{person}{Chuanqi Tao}.}
  \bibinfo{year}{2023}\natexlab{}.
\newblock \showarticletitle{Automated event extraction of {CVE} descriptions}.
\newblock \bibinfo{journal}{\emph{Inf. Softw. Technol.}}  \bibinfo{volume}{158}
  (\bibinfo{year}{2023}), \bibinfo{pages}{107178}.
\newblock


\bibitem[\protect\citeauthoryear{Wen, Chen, Gao, Zhang, Zhang, and Liao}{Wen
  et~al\mbox{.}}{2023}]%
        {DBLP:journals/corr/abs-2302-04675}
\bibfield{author}{\bibinfo{person}{Xin{-}Cheng Wen}, \bibinfo{person}{Yupan
  Chen}, \bibinfo{person}{Cuiyun Gao}, \bibinfo{person}{Hongyu Zhang},
  \bibinfo{person}{Jie~M. Zhang}, {and} \bibinfo{person}{Qing Liao}.}
  \bibinfo{year}{2023}\natexlab{}.
\newblock \showarticletitle{Vulnerability Detection with Graph Simplification
  and Enhanced Graph Representation Learning}. In
  \bibinfo{booktitle}{\emph{Proceedings of the 45th {IEEE/ACM} International
  Conference on Software Engineering ({ICSE})}}. \bibinfo{publisher}{{IEEE}},
  \bibinfo{pages}{2275--2286}.
\newblock


\bibitem[\protect\citeauthoryear{Wu, Shen, Zheng, Lin, Sui, and Semasaba}{Wu
  et~al\mbox{.}}{2023}]%
        {DBLP:journals/kbs/WuSZLSS23}
\bibfield{author}{\bibinfo{person}{Xiaoxue Wu}, \bibinfo{person}{Jinjin Shen},
  \bibinfo{person}{Wei Zheng}, \bibinfo{person}{Lidan Lin},
  \bibinfo{person}{Yulei Sui}, {and} \bibinfo{person}{Abubakar Omari~Abdallah
  Semasaba}.} \bibinfo{year}{2023}\natexlab{}.
\newblock \showarticletitle{RNNtcs: {A} test case selection method for
  Recurrent Neural Networks}.
\newblock \bibinfo{journal}{\emph{Knowl. Based Syst.}}  \bibinfo{volume}{279}
  (\bibinfo{year}{2023}), \bibinfo{pages}{110955}.
\newblock


\bibitem[\protect\citeauthoryear{Yamaguchi, Golde, Arp, and Rieck}{Yamaguchi
  et~al\mbox{.}}{2014}]%
        {DBLP:conf/sp/YamaguchiGAR14}
\bibfield{author}{\bibinfo{person}{Fabian Yamaguchi}, \bibinfo{person}{Nico
  Golde}, \bibinfo{person}{Daniel Arp}, {and} \bibinfo{person}{Konrad Rieck}.}
  \bibinfo{year}{2014}\natexlab{}.
\newblock \showarticletitle{Modeling and Discovering Vulnerabilities with Code
  Property Graphs}. In \bibinfo{booktitle}{\emph{Proceedings of the 35th {IEEE}
  Symposium on Security and Privacy (SP)}}. \bibinfo{publisher}{{IEEE} Computer
  Society}, \bibinfo{pages}{590--604}.
\newblock


\bibitem[\protect\citeauthoryear{Yang, Shi, He, and Lo}{Yang
  et~al\mbox{.}}{2022}]%
        {DBLP:conf/icse/YangSH022}
\bibfield{author}{\bibinfo{person}{Zhou Yang}, \bibinfo{person}{Jieke Shi},
  \bibinfo{person}{Junda He}, {and} \bibinfo{person}{David Lo}.}
  \bibinfo{year}{2022}\natexlab{}.
\newblock \showarticletitle{Natural Attack for Pre-trained Models of Code}. In
  \bibinfo{booktitle}{\emph{Proceedings of the 44th {IEEE/ACM} International
  Conference on Software Engineering ({ICSE})}}. \bibinfo{publisher}{{ACM}},
  \bibinfo{pages}{1482--1493}.
\newblock


\bibitem[\protect\citeauthoryear{Ying, Bourgeois, You, Zitnik, and
  Leskovec}{Ying et~al\mbox{.}}{2019}]%
        {DBLP:conf/nips/YingBYZL19}
\bibfield{author}{\bibinfo{person}{Zhitao Ying}, \bibinfo{person}{Dylan
  Bourgeois}, \bibinfo{person}{Jiaxuan You}, \bibinfo{person}{Marinka Zitnik},
  {and} \bibinfo{person}{Jure Leskovec}.} \bibinfo{year}{2019}\natexlab{}.
\newblock \showarticletitle{GNNExplainer: Generating Explanations for Graph
  Neural Networks}. In \bibinfo{booktitle}{\emph{Proceedings of the 33rd Annual
  Conference on Neural Information Processing Systems (NeurIPS)}}.
  \bibinfo{pages}{9240--9251}.
\newblock


\bibitem[\protect\citeauthoryear{Zeller and Hildebrandt}{Zeller and
  Hildebrandt}{2002}]%
        {DBLP:journals/tse/ZellerH02}
\bibfield{author}{\bibinfo{person}{Andreas Zeller} {and} \bibinfo{person}{Ralf
  Hildebrandt}.} \bibinfo{year}{2002}\natexlab{}.
\newblock \showarticletitle{Simplifying and Isolating Failure-Inducing Input}.
\newblock \bibinfo{journal}{\emph{{IEEE} Trans. Software Eng.}}
  \bibinfo{volume}{28}, \bibinfo{number}{2} (\bibinfo{year}{2002}),
  \bibinfo{pages}{183--200}.
\newblock


\bibitem[\protect\citeauthoryear{Zhang, Wu, and Zhu}{Zhang
  et~al\mbox{.}}{2018}]%
        {DBLP:conf/cvpr/ZhangWZ18a}
\bibfield{author}{\bibinfo{person}{Quanshi Zhang}, \bibinfo{person}{Ying~Nian
  Wu}, {and} \bibinfo{person}{Song{-}Chun Zhu}.}
  \bibinfo{year}{2018}\natexlab{}.
\newblock \showarticletitle{Interpretable Convolutional Neural Networks}. In
  \bibinfo{booktitle}{\emph{Proceedings of the 28th {IEEE} Conference on
  Computer Vision and Pattern Recognition ({CVPR})}}.
  \bibinfo{publisher}{Computer Vision Foundation / {IEEE} Computer Society},
  \bibinfo{pages}{8827--8836}.
\newblock


\bibitem[\protect\citeauthoryear{Zheng, Pujar, Lewis, Buratti, Epstein, Yang,
  Laredo, Morari, and Su}{Zheng et~al\mbox{.}}{2021}]%
        {D2A}
\bibfield{author}{\bibinfo{person}{Yunhui Zheng}, \bibinfo{person}{Saurabh
  Pujar}, \bibinfo{person}{Burn~L. Lewis}, \bibinfo{person}{Luca Buratti},
  \bibinfo{person}{Edward~A. Epstein}, \bibinfo{person}{Bo Yang},
  \bibinfo{person}{Jim Laredo}, \bibinfo{person}{Alessandro Morari}, {and}
  \bibinfo{person}{Zhong Su}.} \bibinfo{year}{2021}\natexlab{}.
\newblock \showarticletitle{{D2A:} {A} Dataset Built for AI-Based Vulnerability
  Detection Methods Using Differential Analysis}. In
  \bibinfo{booktitle}{\emph{Proceedings of the 43rd {IEEE/ACM} International
  Conference on Software Engineering: Software Engineering in Practice
  ({ICSE}-{SEIP})}}. \bibinfo{publisher}{{IEEE}}, \bibinfo{pages}{111--120}.
\newblock


\bibitem[\protect\citeauthoryear{Zhou, Pacheco, Chen, Hu, Xia, Lo, and
  Hassan}{Zhou et~al\mbox{.}}{2023}]%
        {CoLeFunDa}
\bibfield{author}{\bibinfo{person}{Jiayuan Zhou}, \bibinfo{person}{Michael
  Pacheco}, \bibinfo{person}{Jinfu Chen}, \bibinfo{person}{Xing Hu},
  \bibinfo{person}{Xin Xia}, \bibinfo{person}{David Lo}, {and}
  \bibinfo{person}{Ahmed~E. Hassan}.} \bibinfo{year}{2023}\natexlab{}.
\newblock \showarticletitle{CoLeFunDa: Explainable Silent Vulnerability Fix
  Identification}. In \bibinfo{booktitle}{\emph{Proceedings of the 45th
  {IEEE/ACM} International Conference on Software Engineering ({ICSE})}}.
  \bibinfo{publisher}{{IEEE}}.
\newblock


\bibitem[\protect\citeauthoryear{Zhou, Liu, Siow, Du, and Liu}{Zhou
  et~al\mbox{.}}{2019}]%
        {Devign}
\bibfield{author}{\bibinfo{person}{Yaqin Zhou}, \bibinfo{person}{Shangqing
  Liu}, \bibinfo{person}{Jing~Kai Siow}, \bibinfo{person}{Xiaoning Du}, {and}
  \bibinfo{person}{Yang Liu}.} \bibinfo{year}{2019}\natexlab{}.
\newblock \showarticletitle{Devign: Effective Vulnerability Identification by
  Learning Comprehensive Program Semantics via Graph Neural Networks}. In
  \bibinfo{booktitle}{\emph{Proceedings of the 33rd Annual Conference on Neural
  Information Processing Systems (NeurIPS)}}. \bibinfo{pages}{10197--10207}.
\newblock


\bibitem[\protect\citeauthoryear{Zhu, Zhang, Sun, Chen, and Meng}{Zhu
  et~al\mbox{.}}{2023}]%
        {CCzhu}
\bibfield{author}{\bibinfo{person}{Chengcheng Zhu}, \bibinfo{person}{Jiale
  Zhang}, \bibinfo{person}{Xiaobing Sun}, \bibinfo{person}{Bing Chen}, {and}
  \bibinfo{person}{Weizhi Meng}.} \bibinfo{year}{2023}\natexlab{}.
\newblock \showarticletitle{{ADFL:} Defending backdoor attacks in federated
  learning via adversarial distillation}.
\newblock \bibinfo{journal}{\emph{Comput. Secur.}}  \bibinfo{volume}{132}
  (\bibinfo{year}{2023}), \bibinfo{pages}{103366}.
\newblock


\bibitem[\protect\citeauthoryear{Zou, Hu, Li, Wu, Zhao, and Jin}{Zou
  et~al\mbox{.}}{2022}]%
        {mVulPreter}
\bibfield{author}{\bibinfo{person}{Deqing Zou}, \bibinfo{person}{Yutao Hu},
  \bibinfo{person}{Wenke Li}, \bibinfo{person}{Yueming Wu},
  \bibinfo{person}{Haojun Zhao}, {and} \bibinfo{person}{Hai Jin}.}
  \bibinfo{year}{2022}\natexlab{}.
\newblock \showarticletitle{mVulPreter: {A} Multi-Granularity Vulnerability
  Detection System With Interpretations}.
\newblock \bibinfo{journal}{\emph{{IEEE} Trans. Dependable Secur. Comput.}}
  (\bibinfo{year}{2022}).
\newblock


\bibitem[\protect\citeauthoryear{Zou, Zhu, Xu, Li, Jin, and Ye}{Zou
  et~al\mbox{.}}{2021}]%
        {DBLP:journals/tosem/ZouZXLJY21}
\bibfield{author}{\bibinfo{person}{Deqing Zou}, \bibinfo{person}{Yawei Zhu},
  \bibinfo{person}{Shouhuai Xu}, \bibinfo{person}{Zhen Li},
  \bibinfo{person}{Hai Jin}, {and} \bibinfo{person}{Hengkai Ye}.}
  \bibinfo{year}{2021}\natexlab{}.
\newblock \showarticletitle{Interpreting Deep Learning-based Vulnerability
  Detector Predictions Based on Heuristic Searching}.
\newblock \bibinfo{journal}{\emph{{ACM} Trans. Softw. Eng. Methodol.}}
  \bibinfo{volume}{30}, \bibinfo{number}{2} (\bibinfo{year}{2021}),
  \bibinfo{pages}{23:1--23:31}.
\newblock


\end{thebibliography}

\end{document}